\documentclass[aps,prb,reprint,superscriptaddress]{revtex4-1}
\usepackage{amssymb,amsmath,mathptmx}
\usepackage{graphicx}
\usepackage{dcolumn}
\usepackage{bm}
\usepackage{hyperref}
\usepackage{xcolor}
\usepackage[utf8x]{inputenc}
\newcolumntype{P}[1]{>{\centering\arraybackslash}p{#1}}

\begin{document}

\title{The dc-Josephson effect with more than four superconducting
  leads}

\author{R\'egis M\'elin}

\affiliation{Univ. Grenoble-Alpes, CNRS, Grenoble INP\thanks{Institute
    of Engineering Univ. Grenoble Alpes}, Institut NEEL, 38000
  Grenoble, France}

\begin{abstract}
  By definition, the $p$-terminal dc-Josephson current is sensitive to
  the superconducting phase variables of $p$ terminals. In the paper,
  we establish protocol for direct detection of the $p$-terminal
  dc-Josephson effect with $p\ge 3$ in a device containing $N$
  superconducting leads $S_1,\,S_2,\,...,\,S_N$ having the phase
  variables $\varphi_1,\,\varphi_2,\,...,\,\varphi_N$. The calculated
  signal ${\chi}^{(N)}$ can be probed in microwave experiments, and
  it corresponds to the higher-order nonlocal inverse inductance
  obtained from differentiating the current $I_1$ through $S_1$ with
  respect to the remaining $N-2$ independent phase differences
  $\varphi_2-\varphi_N,\,\varphi_3-\varphi_N,\,
  ...,\varphi_{N-1}-\varphi_N$. We find that the values $p\le N-2$ do
  not contribute to ${\chi}^{(N)}$, and that ${\chi}^{(N)}\ne 0$
  implies evidence for the $p=N-1$ or the $p=N$-terminal dc-Josephson
  currents. For $N=4$ superconducting leads, we demonstrate that
  ${\chi}^{(4)}\ne 0$ implies evidence for the $p=3$ or $p=4$
  dc-Josephson effect, irrespective of the $p=2$-terminal dc-Josephson
  current. Thus, we provide a way to demonstrate the dc-Josephson
  effect with more than three terminals ({\it i.e.} with $p\ge 3$) in
  a device containing more than four superconducting leads ({\it i.e.}
  with $N\ge 4$). The predicted ${\chi}^{(4)}$ is ``yes or no'' answer
  to the $p\ge 3$ dc-Josephson effect, {\it i.e.} for $N=4$,
  nonvanishingly small $\chi^{(4)}\ne 0$ implies the $p=3$ or
  $p=4$-terminal dc-Josephson effect and vanishingly small
  $\chi^{(4)}=0$ implies absence of the $p=3$ and $p=4$-terminal
  dc-Josephson effect. The paper can be viewed as generalizing the
  recently considered $\varphi$-junctions in Andreev molecules to
  arbitrary number $N$ of the superconducting leads, and it relies on
  basic properties of the dc-Josephson effect that are not directly
  related to nontrivial topology and Weyl point singularities.
\end{abstract}
\maketitle

\section{Introduction}

\begin{figure*}[htb]
  \begin{minipage}{.7\textwidth}
    \includegraphics[width=.9\textwidth]{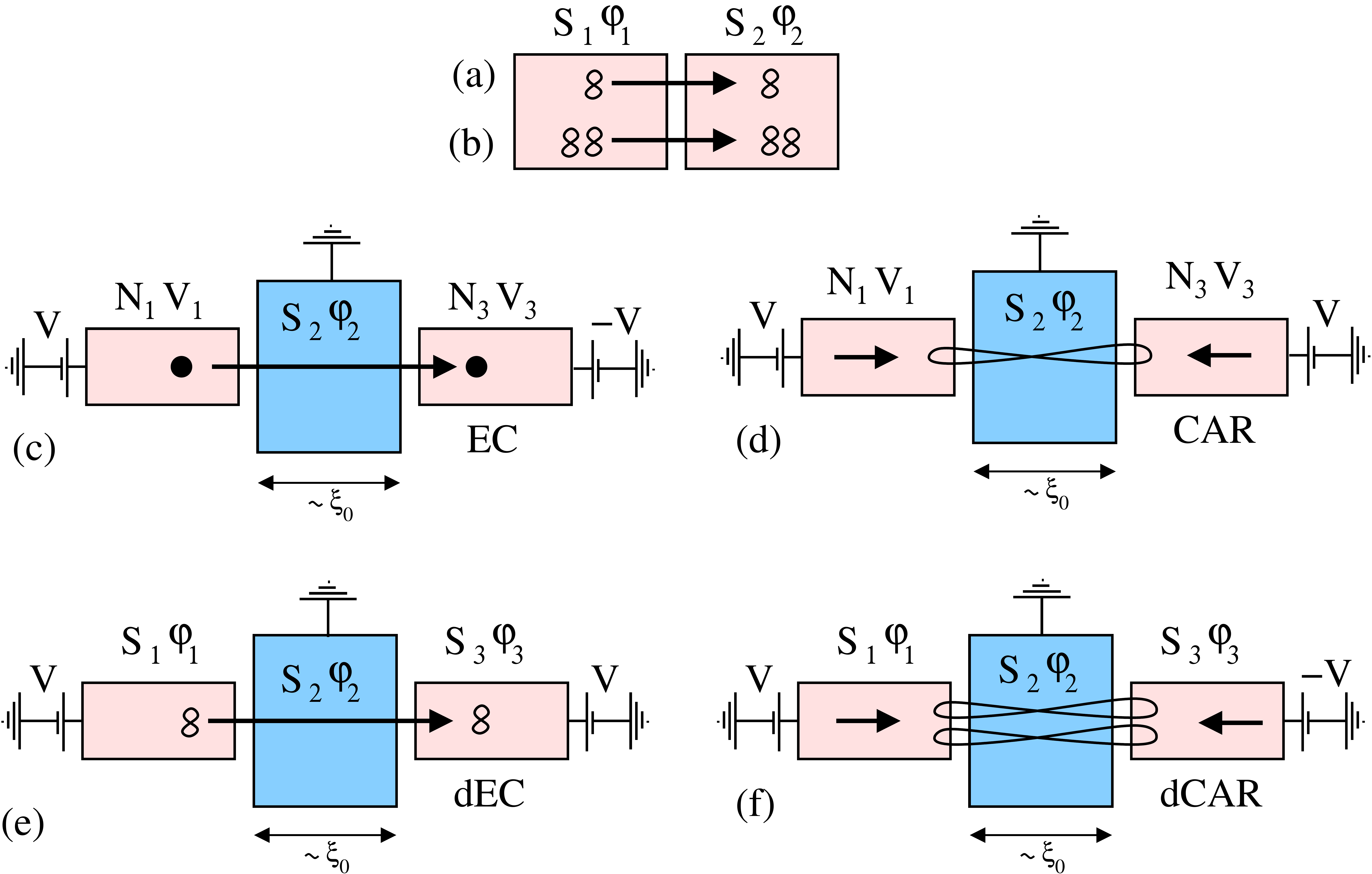}
    \end{minipage}\begin{minipage}{.28\textwidth}
    \caption{Panels a and b show the two-terminal dc-Josephson effect,
      with single Cooper pair tunneling [see
        Eq.~(\ref{eq:IS-2T-tunnel})] and two-Cooper pair tunneling
      [see Eq.~(\ref{eq:IS-2T-gene}) with $n=2$] respectively. Panels
      c and d show double elastic cotunneling (dEC) and double crossed
      Andreev reflection (dCAR) respectively, in a three-terminal
      Josephson junction. Panels e and f show elastic cotunneling (EC)
      and crossed Andreev reflection (CAR) respectively, in a normal
      metal-superconductor-normal metal beam splitter.
    \label{fig:schema}}
  \end{minipage}
\end{figure*}

The two-terminal Josephson junction \cite{Josephson} formed with the
BCS superconductors $S_1$ and $S_2$ reveals physical relevance of the
gauge-invariant difference $\varphi_1 - \varphi_2$ between their
macroscopic phase variables (see
figure~\ref{fig:schema}a). Equilibrium dissipationless supercurrent
\begin{equation}
  \label{eq:IS-2T-tunnel}
  I_S^{(2)}(\varphi_1-\varphi_2) = I_c^{(2)} \sin\left(\varphi_1-\varphi_2
  \right)
\end{equation}
flows across the $S_1$-$S_2$ Josephson weak link \cite{Likharev},
where the superscript ``$(2)$'' in Eq.~(\ref{eq:IS-2T-tunnel}) refers
to the number $N=2$ of the superconducting leads. Anderson and Rowell
\cite{Anderson} experimentally confirmed the prediction
\cite{Josephson} of the dc-Josephson effect.

Eq.~(\ref{eq:IS-2T-tunnel}) is valid for tunnel junctions and the
current-phase relation
\begin{equation}
  \label{eq:IS-2T-gene}
  I_S^{(2)}(\varphi_1-\varphi_2) = \sum_n I_{c,\,n}^{(2)}
  \sin\left[n\left( \varphi_1-\varphi_2 \right)\right]
\end{equation}
with $n$-Cooper pair tunneling holds more generally at arbitrary
interface transparency, see figure~\ref{fig:schema}b for example of
the higher-order two-Cooper pair tunneling with $n=2$ in
Eq.~(\ref{eq:IS-2T-gene}).

{Current flows by Andreev reflection at a normal
  metal-superconductor (NS) interface biased at the voltage $V$
  smaller than the superconducting gap $\Delta$, {\it i.e.}
  $|eV|<\Delta$. Spin-up electron from $N$ is Andreev reflected as a
  spin-down hole while a Cooper pair is transmitted into $S$.}

Nonlocality of Andreev reflection at the scale $R_0\approx \xi_0$ of
the BCS coherence length $\xi_0$ was predicted and experimentally
probed
\cite{theory-CPBS1,theory-CPBS2,theory-CPBS3,theory-CPBS4,theory-CPBS5,theory-CPBS6,theory-CPBS7,theory-CPBS8,theory-CPBS9,exp-CPBS1,exp-CPBS2,exp-CPBS3,exp-CPBS4,exp-CPBS5,exp-CPBS6,exp-CPBS7,exp-CPBS8}
in a $N_1$-$S_2$-$N_3$ three-terminal device since the early 2000's,
see figures~\ref{fig:schema}c and~\ref{fig:schema}d. The normal leads
$N_1$ and $N_3$ are laterally connected to the grounded $S_2$ and the
voltages $V_1$ and $V_3$ are applied on $N_1$ and $N_3$, the
superconducting $S_2$ being grounded at $V_2=0$.

Elastic cotunneling (EC) \cite{theory-CPBS6,theory-CPBS7,theory-CPBS9}
over $R_0\approx \xi_0$ transfers electrons from $N_1$ to $N_3$ across
$S_2$, or from $N_3$ to $N_1$, see figure~\ref{fig:schema}c. Crossed
Andreev reflection (CAR)
\cite{theory-CPBS5,theory-CPBS6,theory-CPBS7,theory-CPBS9} over
$R_0\approx \xi_0$ scatters spin-up electron from $N_1$ as spin-down
hole into $N_2$, leaving a Cooper pair in $S_2$, see
figure~\ref{fig:schema}d. Said differently, two opposite-spin
electrons from $N_1$ and $N_3$ are cooperatively transmitted into the
central $S_2$ and eventually join the condensate.

EC and CAR in the above mentioned $N_1$-$S_2$-$N_3$ three-terminal
device were generalized to dEC \cite{Freyn} (see
figure~\ref{fig:schema}e) and dCAR \cite{Freyn} (see
figure~\ref{fig:schema}f) in a laterally-connected $S_1$-$S_2$-$S_3$
three-terminal Josephson junction with separation $R_0 \approx \xi_0$
between the $S_1$-$S_2$ and $S_2$-$S_3$ interfaces. Namely, double
elastic cotunneling (dEC) \cite{Freyn} on figure~\ref{fig:schema}e
transfers a Cooper pair from $S_1$ to $S_3$ across $S_2$, or from
$S_3$ to $S_1$. Double crossed Andreev reflection (dCAR) \cite{Freyn}
on figure~\ref{fig:schema}f takes a Cooper pair from $S_1$ and another
one from $S_2$. The two pairs from $S_1$ and $S_3$ exchange partners,
yielding four-fermion state, {\it i.e.}  the so-called quartet of
electrons
\cite{Freyn,Melin-EPJB,QUARTETS1,Rech,QUARTETS2,QUARTETS3,QUARTETS4,QUARTETS5,QUARTETS6,QUARTETS7}. The
quartet eventually dissociates as two ``outgoing'' Cooper pairs
joining the condensate of $S_2$.

Voltage biasing the $S_1$-$S_2$-$S_3$ three-terminal Josephson
junction \cite{Freyn,Melin-EPJB} on figures~\ref{fig:schema}e
and~\ref{fig:schema}f at the voltage differences $V_1-V_2$ and
$V_3-V_2$ is a possibility to reveal \cite{Lefloch,Heiblum,Kim} dCAR
and the quartets as emergence of the $V_1+V_3=0$ dc-Josephson
resonance line, if one of the elements of the differential conductance
matrix is plotted in the $(V_1,V_3)$ voltage plane while the
``central'' $S_2$ is grounded at the reference voltage $V_2=0$. Other
dc-Josephson resonance lines were predicted\cite{Freyn,Melin-EPJB} and
observed\cite{Lefloch,Heiblum,Kim}, such as $V_1=V_3$ due to
dEC. {Other experiments
  \cite{multiterminal-exp1,multiterminal-exp2,multiterminal-exp3,multiterminal-exp4,multiterminal-exp5}
  did not report the quartet dc-Josephson anomaly, which is maybe a
  matter of the materials and configurations of the devices. In
  addition, the recent Ref.~\onlinecite{multiterminal-exp6} reports
  progress in the fabrication of four-terminal devices.}

Considering more generally $p_1$ Cooper pairs from $S_1$ and $p_3$
from $S_3$ yields the energy $E_{initial}= 2 p_1 V_1 + 2 p_3 V_3$ of
the ``initial state'', and $E_{final} = 2\left(p_1 + p_3\right) V_2$
for the final state, with $E_{final}\equiv 0$ because $V_2=0$ for the
grounded $S_2$. Energy conservation $E_{initial}=E_{final}$ implies
the dc-Josephson resonance line at
\begin{equation}
  \label{eq:res-V}
  p_1 V_1 + p_3 V_3 = 0
\end{equation}
which is sustained by the static dc-phase variable
\begin{eqnarray}
  \label{eq:phi-p1-q1-t}
  \varphi_{p_1,q_1}&=& p_1 \left[\varphi_1(t)-\varphi_2(t)\right]
  + p_3 \left[\varphi_3(t)-\varphi_2(t)\right]\\
  &=&p_1 \left[\varphi_1-\varphi_2\right]
  + p_3 \left[\varphi_3-\varphi_2\right]
    \label{eq:phi-p1-q1-0}
.
\end{eqnarray}
The following Josephson relations \cite{Josephson}
\begin{eqnarray}
  \varphi_1(t)&=&\varphi_1 + \frac {2eV_1}{\hbar}
  t\\ \varphi_2(t)&=&\varphi_2\\ \varphi_3(t)&=&\varphi_3 + \frac
  {2eV_3}{\hbar} t
\end{eqnarray}
for the superconducting phase variables $\varphi_1(t),\,\varphi_2(t)$
and $\varphi_3(t)$ as a function of the time~$t$ were combined to
Eq.~(\ref{eq:res-V}) in order to deduce Eq.~(\ref{eq:phi-p1-q1-0})
from Eq.~(\ref{eq:phi-p1-q1-t}).  Then, the multipair supercurrent
generalizing the quartets \cite{Freyn,Melin-EPJB} takes the form
\begin{equation}
  \label{eq:IJ-3T-gene-ori}
  I_{p_1,p_3}^{(3)}= I_{c,p_1,p_3} \sin\left[
    p_1\left(\varphi_1-\varphi_2\right) +p_3
    \left(\varphi_3-\varphi_2\right) \right] .
\end{equation}

However, Eq.~(\ref{eq:IJ-3T-gene-ori}) also holds at equilibrium, {\it
  i.e.} if all superconducting leads $S_1$, $S_2$ and $S_3$ are
grounded at $V_1-V_2=V_3-V_2=0$, and biased at the phase differences
$\varphi_1-\varphi_2$ and $\varphi_3-\varphi_2$. The present paper
focuses on the equilibrium multiterminal dc-Josephson effect where all
leads are grounded.

Nonlocality of the dc-Josephson effect can be understood in different
ways:

(i) Subgap propagation over the zero-energy BCS coherence length
$R_0\approx \xi_0$ across the ``central'' $S_2$ in a $S_1$-$S_2$-$S_3$
three-terminal Josephson junction made with the $S_1$-$S_2$ and
$S_2$-$S_3$ lateral contacts separated by $R_0$, see the above
discussion.

(ii) Emergence of dc-Josephson current controlled by the phase of
three or more superconducting leads, see Eq.~(\ref{eq:IJ-3T-gene-ori})
with $p_1 \ne 0$ and $p_3\ne 0$.

\begin{figure*}[htb]
  \begin{minipage}{.7\textwidth}
    \includegraphics[width=.9\textwidth]{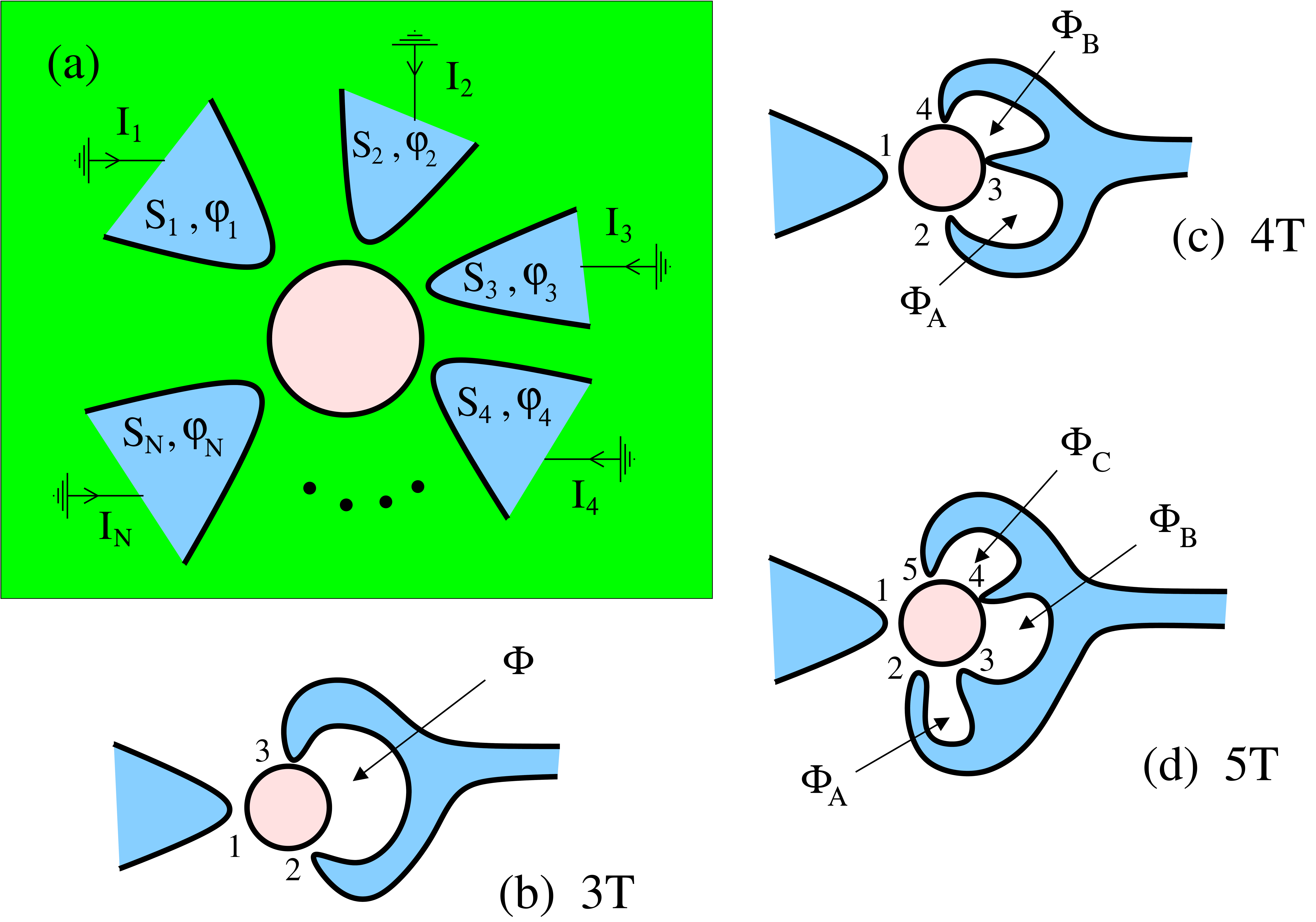}
    \end{minipage}\begin{minipage}{.28\textwidth}
  \caption{The highlighted panel a shows the $N$-terminal Josephson
    junction considered in the paper, where the $N$ grounded
    superconducting leads $S_1,\,\,S_2,\,...,\,S_N$ with
    superconducting phase variables $\varphi_1,\,...,\,\varphi_N$ are
    connected to a small nonsuperconducting region. The supercurrents
    through the leads $S_1,\,\,S_2,\,...,\,S_N$ are denoted by
    $I_1,\,I_2,\,...,\,I_N$, with $\sum_{\alpha=1}^N
    I_\alpha=0$. Panels b-d show the devices with $N=3$~(b),
    $N=4$~(c), and $N=5$~(d). 
    \label{fig:N-terminal-junction}}
  \end{minipage}
\end{figure*}

The considered devices involve $N$ superconductors connected by
single-channel weak links to the nonsuperconducting ``central''
region, see figure~\ref{fig:N-terminal-junction}a. The above item (i)
for nonlocality over $R_0 \approx \xi_0$ is thus not directly relevant
to the present work. According to the above item (ii), we evaluate the
contribution of the $p$-terminal Josephson effect in a device
containing $N$ superconducting leads. Namely, we evaluate the
sensitivity of the current $I_1$ through lead $S_1$ on the $p$
superconducting phase variables
$\varphi_{a_1},\,\varphi_{a_2},\,...,\varphi_{a_p}$ where $1\le
a_1<a_2<...<a_p\le N$. Then, we define $\chi^{(N)}$ as the partial
derivative of $I_1$ with respect to the $N-2$ phase differences
$\varphi_2-\varphi_N$, $\varphi_3-\varphi_N$, ...,
$\varphi_{N-1}-\varphi_N$. We show that the $p$-terminal dc-Josephson
effect with $p\le N-2$ does not contribute to $\chi^{(N)}$ while $p=N$
and $p=N-1$ yield nonvanishingly small contribution to
$\chi^{(N)}$. Thus, microwave experiments can reveal experimental
evidence for $\chi^{(4)}\ne 0$ with $N=4$, which implies the $p=3$ or
the $p=4$ three or four-terminal dc-Josephson effects, whatever the
value of the $p=2$ dc-Josephson supercurrent. Thus, the following
paper demonstrates the possibility of directly testing the
multiterminal dc-Josephson effect with $p\ge 3$ in microwave
experiments.

The paper is organized as follows. Section~\ref{sec:connection}
establishes connection to known
results. Section~\ref{sec:p-N-terminals} presents the model and the
$p$-terminal Josephson current in a device containing $N$
superconducting leads. Section~\ref{sec:N=34examples} provides
examples with $N=3$ and $N=4$ superconducting leads.
Section~\ref{sec:general-N} generalizes the theory to arbitrary number
$N$ of the superconducting leads. Numerical results for $N=4$
superconducting leads are presented in section~\ref{sec:num}. Summary
and final remarks are provided in section~\ref{sec:conclusions}.

\section{Connection with recent results}
\label{sec:connection}

{We first note that the field of multiterminal Josephson junctions
  \cite{Freyn,Melin-EPJB,QUARTETS1,Rech,QUARTETS2,QUARTETS3,QUARTETS4,QUARTETS5,QUARTETS6,QUARTETS7}
  has recently been enriched with the proposal of nontrivial topology
  \cite{Nazarov1,Nazarov2,topo0,topo1,topo2,topo3,topo4,topo5,Levchenko1,Levchenko2,Berry}
  and topology in the time-periodic Floquet dynamics
  \cite{Feinberg1,Feinberg2,topo1-plus-Floquet}. The present paper is
  not directly related to those works on topology, in the sense that
  we consider basic properties of the dc-Josephson, which generally
  hold in the absence of Weyl-point singularities.}

This section provides connection to known results on the multiterminal
dc-Josephson interferometers discussed in
Refs.~\onlinecite{Pillet1,Pillet2,Benzoni,Rech}.

The recent Refs.~\onlinecite{Pillet1,Pillet2,Benzoni} considered an
``Andreev molecule'' in the laterally-connected $S_1$-$S_2$-$S_3$
three-terminal Josephson junction under equilibrium voltage biasing
conditions, {\it i.e.}  the three superconducting leads $S_1,\,S_2$
and $S_3$ are grounded at
$V_1-V_2=V_3-V_2=0$. Refs.~\onlinecite{Pillet1,Pillet2,Benzoni}
characterize the supercurrent $I_1(\varphi_3)$ through lead $S_1$ as a
function of the phase $\varphi_3\ne 0$ on $S_3$ which is not directly
connected to $S_1$, while $\varphi_2=0$ is the reference, and it is
assumed in addition that $\varphi_1=0$.

This nonlocal response \cite{Pillet1,Pillet2,Benzoni} of the
supercurrent $I_1$ through $S_1$ to the phase $\varphi_3$ on $S_3$
originates from the above mentioned dEC and dCAR \cite{Freyn}. In the
limit of tunnel contacts, the dEC and dCAR Josephson currents are given by
\cite{Freyn}
\begin{eqnarray}
I_{dEC}^{(3)}&=&I_{c,\,dEC}^{(3)} \sin(\varphi_1-\varphi_3) ,
  \label{eq:dEC}\\
  \label{eq:dCAR}
  I_{dCAR}^{(3)}&=&I_{c,\,dCAR}^{(3)}
  \sin(\varphi_1+\varphi_3-2\varphi_2)
  ,
\end{eqnarray}
where ``$(3)$'' in the superscript refers to the number $N=3$ of the
superconducting leads. Considering $\varphi_1=\varphi_2=0$ and
$\varphi_3\ne 0$ yields the dc-Josephson current $I_1$ through $S_1$:
\begin{equation}
    \label{eq:phi}
  I_1^{(3)}(0)=I_{dEC}^{(3)}+I_{dCAR}^{(3)}
  =\left(-I_{c,\,dEC}^{(3)}+I_{c,\,dCAR}^{(3)}\right)\sin\varphi_3
  .\end{equation} In an intuitive two-terminal picture, a
$\varphi$-junction is obtained \cite{Pillet1,Pillet2,Benzoni}: the
supercurrent $I_1^{(3)}(\varphi_1-\varphi_2)$ plotted as a function of
$\varphi_1-\varphi_2$ is shifted by arbitrary phase which is
controlled by $\varphi_3-\varphi_2\ne 0$. Two limiting cases are
considered:

(a) {\it In the tunnel limit}, the dEC and dCAR current-phase
relations are $0$- and $\pi$-shifted respectively
\cite{Freyn,Melin-EPJB}. Then, $I_{c,\,dCAR}^{(3)}<0$ and
$-I_{c,\,dEC}^{(3)}<0$ are both negative, yielding the same negative
sign $-I_{c,\,dEC}^{(3)}+I_{c,\,dCAR}^{(3)}<0$ as if
$-I_{c,\,dEC}^{(3)}$ was alone. In this item~(a), we assume that the
contacts have linear dimension which is large compared to the Fermi
wave-length~$\lambda_F$ and small compared to the superconducting
coherence length~$\xi_0$.

(b) {\it The limit $R_0\ll\xi_0$ of small separation between the
  contacts} necessarily produces $I_{c,\,dCAR}^{(3)}\simeq 0$, because
the assumption $R_0\ll\xi_0$ implies equivalence to the two-terminal
$S_1$-$S_3$. Then, the sign of
$-I_{c,\,dEC}^{(3)}+I_{c,\,dCAR}^{(3)}\simeq -I_{c,\,dEC}^{(3)}<0$ is
necessarily negative in the absence of dCAR.

The conditions for deducing ``evidence for dCAR'' from positive
\begin{equation}
  \label{eq:positive}
  -I_{c,\,dEC}^{(3)}+I_{c,\,dCAR}^{(3)}>0
\end{equation}
have thus not been elucidated at present time: The above arguments
show that Eq.~(\ref{eq:positive}) is parameter-dependent. In the
following paper, we propose another possibility for directly probing
the $p$-terminal dc-Josephson effect with $p\ge 3$.

Considering now the $(S_1,\,S_2,\,S_3,\,S_4)$ four-terminal Josephson
junction with $N=4$, we recently proposed \cite{Rech} a
superconducting quantum interference device (SQUID \cite{SQUID})
containing two loops making four contacts on a double quantum dot
defined in a semiconducting nanowire or a carbon nanotube. It was
shown in Ref.~\onlinecite{Rech} that dEC and dCAR can be distinguished
as different peaks in the Fourier transform of the critical current
with respect to the magnetic field. Thus, this double SQUID
\cite{Rech} successfully proposes a way to provide evidence for dCAR
and for ``the three-terminal dc-Josephson effect''.

The device on figure~\ref{fig:N-terminal-junction}c contains the four
superconducting leads $S_1,\,S_2,\,S_3,\,S_4$ which are biased at the
three independent phases differences $\varphi_1-\varphi_4$,
$\varphi_2-\varphi_4$ and $\varphi_3-\varphi_4$. As it is mentioned
above, the superconducting leads are grounded at $V_1-V_4=0$,
$V_2-V_4=0$ and $V_3-V_4=0$.  In addition, the two loops are pierced
by the magnetic field fluxes $\Phi_A$ and $\Phi_B$. Compared to
Ref.~\onlinecite{Rech}, we complementary propose microwave detection
of the second-order nonlocal inverse inductance
\begin{eqnarray}
  \label{eq:S-(4)-gene}
  {\chi}^{(4)}&\equiv& {\chi}^{(4)} \left(\varphi_1-\varphi_4,
  \varphi_2-\varphi_4, \varphi_3-\varphi_4\right)\\ &=&
  \frac{\partial^2 I_1^{(4)} \left(\varphi_1-\varphi_4,
    \varphi_2-\varphi_4, \varphi_3-\varphi_4\right)}{\partial
    \left(\varphi_2-\varphi_4\right)
    \partial\left(\varphi_3-\varphi_4\right)} ,
  \label{eq:S-(4)-gene-bis}
\end{eqnarray}
which is also evaluated as a function of $\varphi_1-\varphi_4$
for $\varphi_2-\varphi_4=0$ and $\varphi_3-\varphi_4=0$:
\begin{eqnarray}
  \label{eq:S-(4)}
&&{\chi}^{(4)(0)}\equiv {\chi}^{(4)(0)}\left(\varphi_1-\varphi_4\right)\\&=&
  \left.\frac{\partial^2 I_1^{(4)}\left(\varphi_1-\varphi_4,
    \varphi_2-\varphi_4, \varphi_3-\varphi_4\right) }{\partial
    \left(\varphi_2-\varphi_4\right)
    \partial\left(\varphi_3-\varphi_4\right)}
  \right|_{\begin{array}{c}\varphi_2-\varphi_4=0\\\varphi_3-\varphi_4=0
  \end{array}} .
  \label{eq:S-(4)-bis}
\end{eqnarray}
The superscript ``$(4)$'' in
Eqs.~(\ref{eq:S-(4)-gene})-(\ref{eq:S-(4)-gene-bis}) and
Eqs.~(\ref{eq:S-(4)})-(\ref{eq:S-(4)-bis}) refers to the number $N=4$
of the superconducting leads, see the device on
figure~\ref{fig:N-terminal-junction}c. {In experiments,
  Eqs.~(\ref{eq:S-(4)-gene})-(\ref{eq:S-(4)-bis}) can be detected with
  microwaves.}

\begin{table*}[ht]
\begin{tabular}{ |P{2.2cm}||P{2.2cm}||P{2.2cm}|P{2.2cm}|P{2cm}|P{1.6cm}|P{1.6cm}|P{1.6cm}|}
  \hline $N$& Figure number & $p=2$ & $p=3$ & $p=4$ & ... & $p=N-1$ &
  $p=N$ \\ \hline $2$ & &
  Eqs.~(\ref{eq:IS-2T-tunnel})-(\ref{eq:IS-2T-gene}) & & &... & &
  \\ \hline $3$ & Fig.~\ref{fig:N-terminal-junction}b &
  Eqs.~(\ref{eq:I-panelb-debut-1-ori})-(\ref{eq:I-panelb-debut-1-ori-2})
  & Eq.~(\ref{eq:I-panelb-debut-1-ori-fin}) & & ...& & \\ \hline $4$
  &Fig.~\ref{fig:N-terminal-junction}c
  &Eqs.~(\ref{eq:I-panelb-debut-1})-(\ref{eq:BB1})&
  Eqs.~(\ref{eq:box-page-6})-(\ref{eq:AA2})
  &Eq.~(\ref{eq:star-page-7}) & ...& & \\ &
  &&Eq.~(\ref{eq:TOTO1})&Eq.~(\ref{eq:TOTO1-fin})&...&&\\ &&&Eq.~(\ref{eq:TOTO1-0})&Eq.~(\ref{eq:TOTO1-0})&...&&\\ &Fig.~\ref{fig:cartoon_for_m}&Fig.~\ref{fig:cartoon_for_m}-a1&Fig.~\ref{fig:cartoon_for_m}-a2&
  Fig.~\ref{fig:cartoon_for_m}-a3&...&&\\ &Fig.~\ref{fig:cartoon_for_dressing}&Fig.~\ref{fig:cartoon_for_dressing}-a1&Fig.~\ref{fig:cartoon_for_dressing}-a2&
  &...&&\\ \hline ... & ... &... & ... & ... & ... & ... &
  ... \\ \hline $N=2q$ &Fig.~\ref{fig:N-terminal-junction}a& & & &...
  &Eq.~(\ref{eq:IprimeS-2-even-2-B})
  &Eq.~(\ref{eq:IprimeS-2-even-2-A})\\ \hline $N=2q'+1$
  &Fig.~\ref{fig:N-terminal-junction}a& & & &...
  &Eq.~(\ref{eq:IprimeS-2-odd-2-B-x})
  &Eq.~(\ref{eq:IprimeS-2-odd-2-A})\\ \hline
\end{tabular}
\caption{The table summarizes the values of $p$ and $N$ that are
  considered in the paper.
    \label{table} 
}
\end{table*}

In the paper, we demonstrate that

``{\it ${\chi}^{(4)}\left(\varphi_1-\varphi_4,\Phi_A,\Phi_B\right)=0$ or
  ${\chi}^{(4)(0)}\left(\varphi_1-\varphi_4\right)=0$}''

generically implies

``{\it Absence of the three- and four-terminal dc-Josephson effect, 
i.e.  $I_{c,p_1,p_3}^{(3)}=0$ in Eq.~(\ref{eq:IJ-3T-gene-ori}) for all
$p_1,\,p_3$, and $I_{c,p_1,p_2,p_3}^{(4)}=0$ for all
$p_1,\,p_2,\,p_3$}''.

The above mentioned four-terminal critical current
\cite{QUARTETS6,Kim} $I_{c,p_1,p_2,p_3}^{(4)}$ corresponds to
\begin{eqnarray}
  \label{eq:IJ-3T-gene}
  && I_{p_1,p_2,p_3}^{(4)}
  \left(\varphi_1-\varphi_4,\varphi_2-\varphi_4,
  \varphi_3-\varphi_4\right)\\  \nonumber
  &=& I_{c,p_1,p_2,p_3}^{(4)} \sin\left[
    p_1\left(\varphi_1-\varphi_4\right) +p_2
    \left(\varphi_2-\varphi_4\right)\right.\\&&\left. +p_3
    \left(\varphi_3-\varphi_4\right) \right] .  \nonumber
\end{eqnarray}

Conversely, we deduce a second logical link:

``{\it Nonvanishingly small signal ${\chi}^{(4)}\left(\varphi_1-\varphi_4,\Phi_A,\Phi_B\right)\ne 0$ or
  ${\chi}^{(4)(0)}\left(\varphi_1-\varphi_4\right)\ne 0$}''

generically implies

``{\it Evidence for the three- or four-terminal dc-Josephson effect, i.e.  
$I_{c,p_1,p_3}^{(3)}\ne 0$ for some $p_1,\,p_3$ or
  $I_{c,p_1,p_2,p_3}^{(4)}\ne 0$ for some $p_1,\,p_2,\,p_3$}''.

The two logical links are independent on the value of the two-terminal
dc-Josephson current, {\it i.e.} $I_{c,\,n}^{(2)}$ in
Eq.~(\ref{eq:IS-2T-gene}) can take the arbitrary values
$I_{c,\,n}^{(2)}\ne 0$. In addition, we generalize the theory to an
arbitrary number $N$ of the superconducting leads.

\section{$p$-terminal Josephson current
  with $N$ superconducting leads}
\label{sec:p-N-terminals}
In this section, we establish an expression for the $p$-terminal
dc-Josephson current with $N\ge p$ superconducting leads. The Table
summarizes the different values of $p$ and $N$ that are considered in
the paper.  Subsection~\ref{sec:ABS-energies} presents the expression
of the Andreev Bound States (ABS) energies. Subsection~\ref{sec:p-ter}
provides the general expression of the $p$-terminal Josephson
current. Subsection~\ref{sec:dot} presents the example of a
single-level quantum dot in the infinite-gap limit.

\subsection{Expression of the supercurrent carried by the Andreev Bound
  States}
\label{sec:ABS-energies}

In this subsection, we evaluate the Fourier transform of the ABS
energies with respect to all superconducting phase variables
$\varphi_1$,\,..., $\varphi_N$ of the $N$ superconducting leads
$S_1,\,...,\,S_N$ connected to small nonsuperconducting region.

The ABS energies $E^{\{\varphi\}}_{ABS, \lambda}\left(
\varphi_1,\,...,\,\varphi_N\right)$ depend on all values of the
$\varphi_n-\varphi_N$ (where $n=1,\,...,\,N-1$ labels the
superconducting leads and $\lambda=1,\,...,\,N_{ABS}$ is used for the
negative-energy ABS). Taking the Fourier transform leads to
\begin{widetext}
\begin{eqnarray}
  \label{eq:E-ABS1}
E^{\{\varphi\}}_{ABS,\lambda}\left( \varphi_1,\,\varphi_2,
...,\,\varphi_N\right)&=&\sum_{m_1} \sum_{m_2} ... \sum_{m_N}
E^{\{m\}}_{ABS,\lambda}\left[
  m_1,\,m_2,\,...,\,m_N\right]\times\\&& \exp\left(i\left(m_1
\varphi_1 + m_2\varphi_2 + .... + m_N\varphi_N\right)\right)
\delta\left(m_1+m_2+...+m_N\right)
,
\nonumber
\end{eqnarray}
\end{widetext}
where the constraint
\begin{equation}
  \label{eq:constraint}
  m_1+m_2+...+m_N=0
\end{equation}
originates from gauge invariance, {\it i.e.} any of the
superconducting phases $\varphi_{n_{ref}}$ can be chosen as the
reference, leaving $N-1$ independent gauge-invariant phase differences
$\varphi_n-\varphi_{n_{ref}}$ with $n\ne n_{ref}$. For instance,
choosing $n_{ref}=N$ leads to
\begin{eqnarray}
\label{eq:gauge-invariance}
  &&m_1 \varphi_1 + m_2\varphi_2 + .... + m_N\varphi_N\\ &=&m_1\left(
\varphi_1-\varphi_N\right) + m_2\left(\varphi_2 -\varphi_N\right) +
...  \nonumber\\&&+ m_{N-1}\left(\varphi_{N-1} -\varphi_N\right),
\nonumber
\end{eqnarray}
where Eq.~(\ref{eq:constraint}) and Eq.~(\ref{eq:gauge-invariance})
are equivalently used. The variables $m_1,\,m_2,\,...,\,m_N$ are
conjugate to the superconducting phases $\varphi_1,
\varphi_2,\,...,\,\varphi_N$, {\it i.e.} they are interpreted as the
algebraic number of Cooper pairs transmitted into $S_1,\,S_2,\,...,\,S_N$.

The zero-temperature dc-Josephson current through the lead $S_{N_0}$ is
given by
\begin{widetext}
\begin{eqnarray}
  &&I_{N_0}^{S,\,\left(N\right)}\left(\varphi_{1},\,...,\,\varphi_{N}\right)
  \label{eq:IS-N0-a1}
  =-\frac{e}{\hbar} \frac{\partial}{\partial \varphi_{N_0}}
  \sum_\lambda E^{\{\varphi\}}_{ABS,\lambda}\left(
  \varphi_1,\,\varphi_2,\,...,\,\varphi_N\right)\\ &=&
  \sum_{m_1} \sum_{m_2} ... \sum_{m_N}
  I_{c,\,n_0}^{\left(N\right)}\left[m_1,\,...,\,m_N\right]
   \sin\left(m_1 \varphi_1 +
   m_2\varphi_2 + .... + m_N\varphi_N\right)
   \delta\left(m_1+m_2+...+m_N\right)
  \label{eq:IS-N0-a2}
,
\end{eqnarray}
\end{widetext}
where
\begin{equation}
  \label{eq:Ic-N0}
  I_{c,\,n_0}^{\left(N\right)}\left[m_1,\,...,\,m_N\right]=-\frac{e}{\hbar} m_{N_0}
  E^{\{m\}}_{ABS,\lambda}\left[m_1,\,m_2,\,...,\,m_N\right]
  .
\end{equation}
  
\begin{figure*}[htb]
  \includegraphics[width=.7\textwidth]{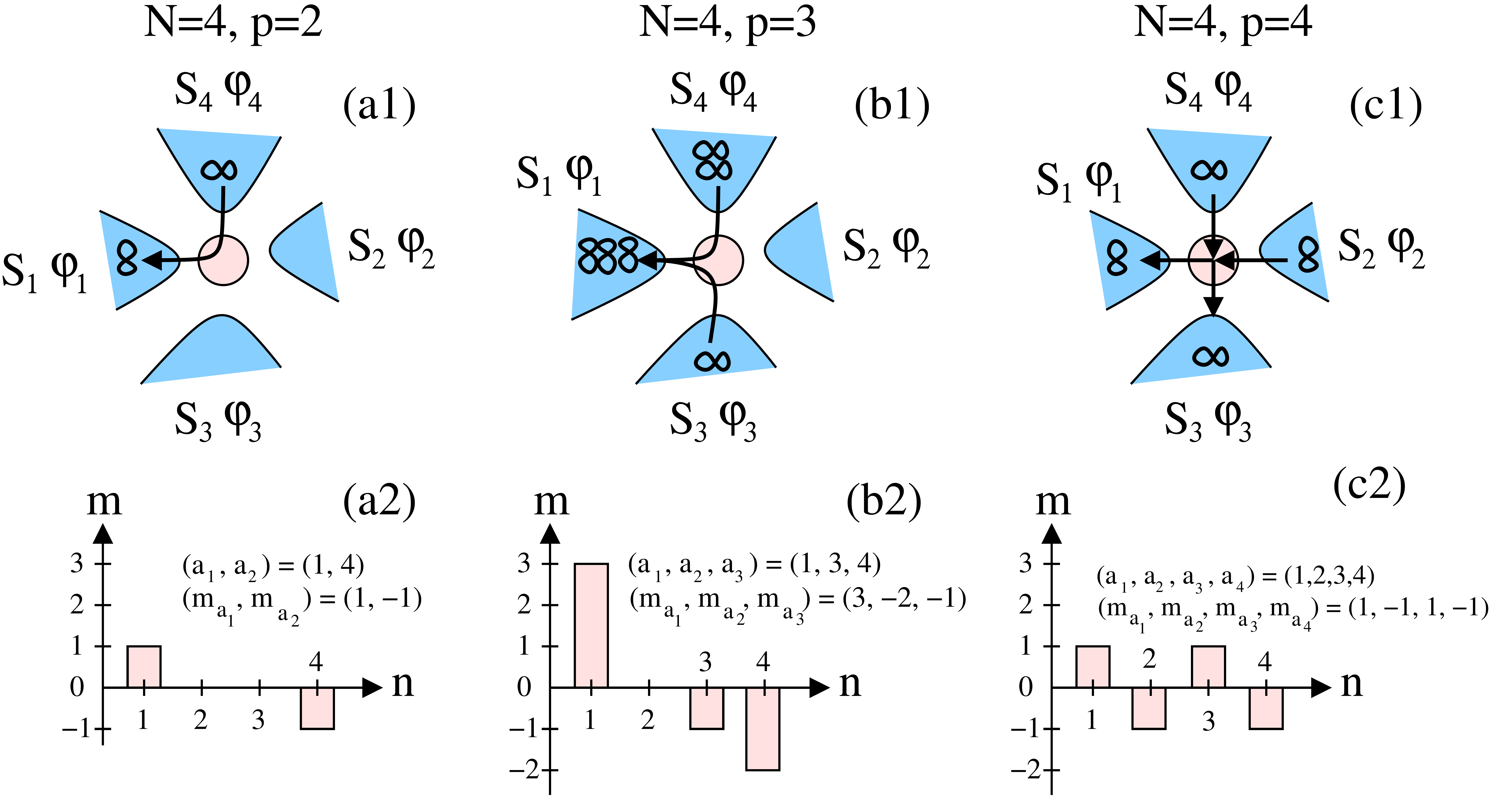}
  \caption{Panels a1, b1 and c1 show the ``incoming'' and ``outgoing''
    Cooper pairs for the $p=2,\,3,\,4$-terminal dc-Josephson processes
    respectively, with a total of $N=4$ superconducting leads. The
    corresponding values of $a_\alpha$ and $m_{a_\alpha}$ are shown on
    panels a2, b2 and c2. The integer $a_\alpha \in\{1,\,...,\,n\}$ is
    shown on the $x$-axis of panels a2, b2 and c2 corresponding to the
    parameters of panels a1, b1 and c1 respectively. The variable
    $m\equiv m_\alpha$ on the $y$-axis of panels a2, b2 and c2 is the
    algebraic number of Cooper pairs transmitted in each
    superconducting lead.
      \label{fig:cartoon_for_m} }
\end{figure*}
\begin{figure*}[htb]
  \begin{minipage}{.66\textwidth}
    \includegraphics[width=.7\textwidth]{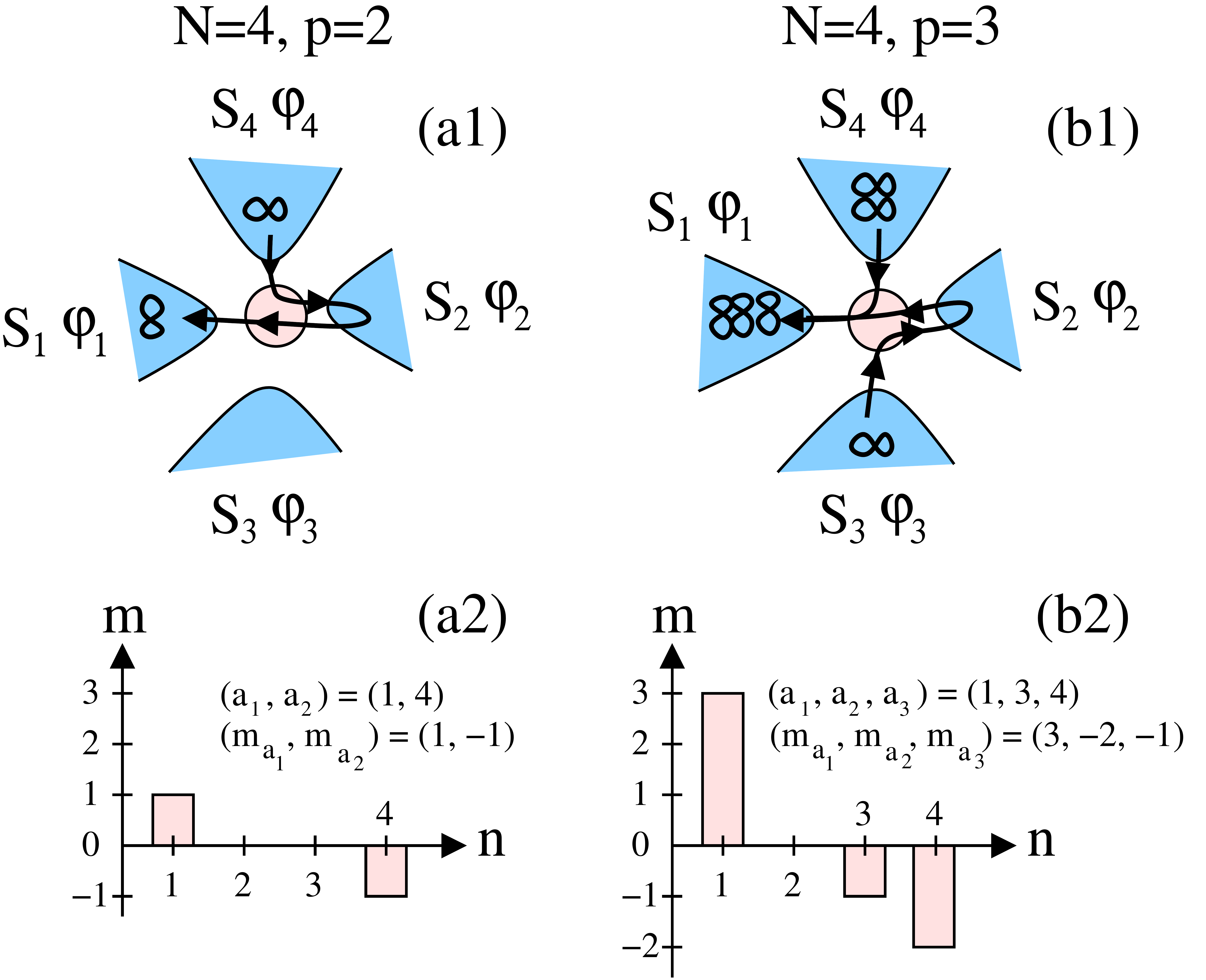}
    \end{minipage}\begin{minipage}{.3\textwidth}
  \caption{Panels a1 and a2 are obtained by ``dressing'' the $p=2$ and
    $p=3$-terminal dc-Josephson current processes on
    figures~\ref{fig:cartoon_for_m}-a1 and \ref{fig:cartoon_for_m}-a2
    by an ``excursion'' in and out from one of the superconducting
    leads. Thus, the $p=2$ and $p=3$-terminal dc-Josephson processes
    on panel a1 and b1 involve three and four terminal
    respectively. They receive the same $a_\alpha$ and $m_{a_\alpha}$
    values as the ``bare'' diagrams on
    figures~\ref{fig:cartoon_for_m}-a1 and
    \ref{fig:cartoon_for_m}-a2. 
    \label{fig:cartoon_for_dressing} }
  \end{minipage}
\end{figure*}

\begin{figure*}[htb]
    \includegraphics[width=.7\textwidth]{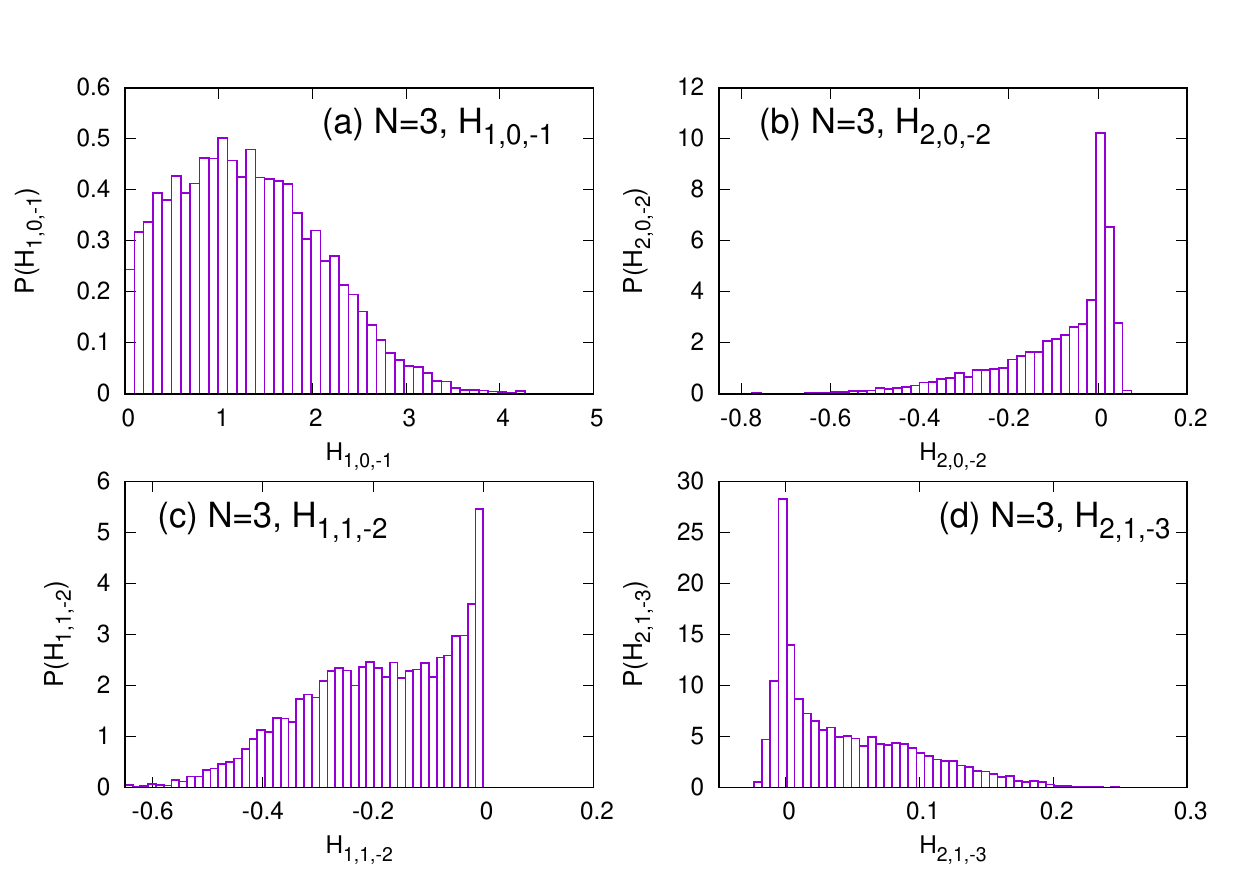}
    \caption{The figure shows the distribution
      $P(H_{n_1,\,\,n_2,\,\,n_3})$ for a quantum dot connected to $N=3$
      superconducting leads, where $H_{n_1,\,\,n_2,\,\,n_3}$ is given by
      Eq.~(\ref{eq:harmo-n}) with $N=p=3$. Panels a, b, c and d
      correspond to $(n_1,\,\,n_2,\,\,n_3)=(1,\,0,\,-1)$, $(2,\,0,\,-2)$,
      $(1,\,1,\,-2)$ and $(2,\,1,\,-3)$ respectively. The coupling
      parameters $\Gamma$s entering
      Eqs.~(\ref{eq:H-infinite})-(\ref{eq:E-ABS-gap-infini}) are
      chosen at random in a Gaussian distribution centered at
      $\overline{\Gamma}=1$ and with root-mean-square $\delta
      \Gamma=0.5$. The negative values of the $\Gamma$s are rejected.
    \label{fig:figure-histos-N3}}
\end{figure*}

\begin{figure*}[htb]
    \includegraphics[width=.7\textwidth]{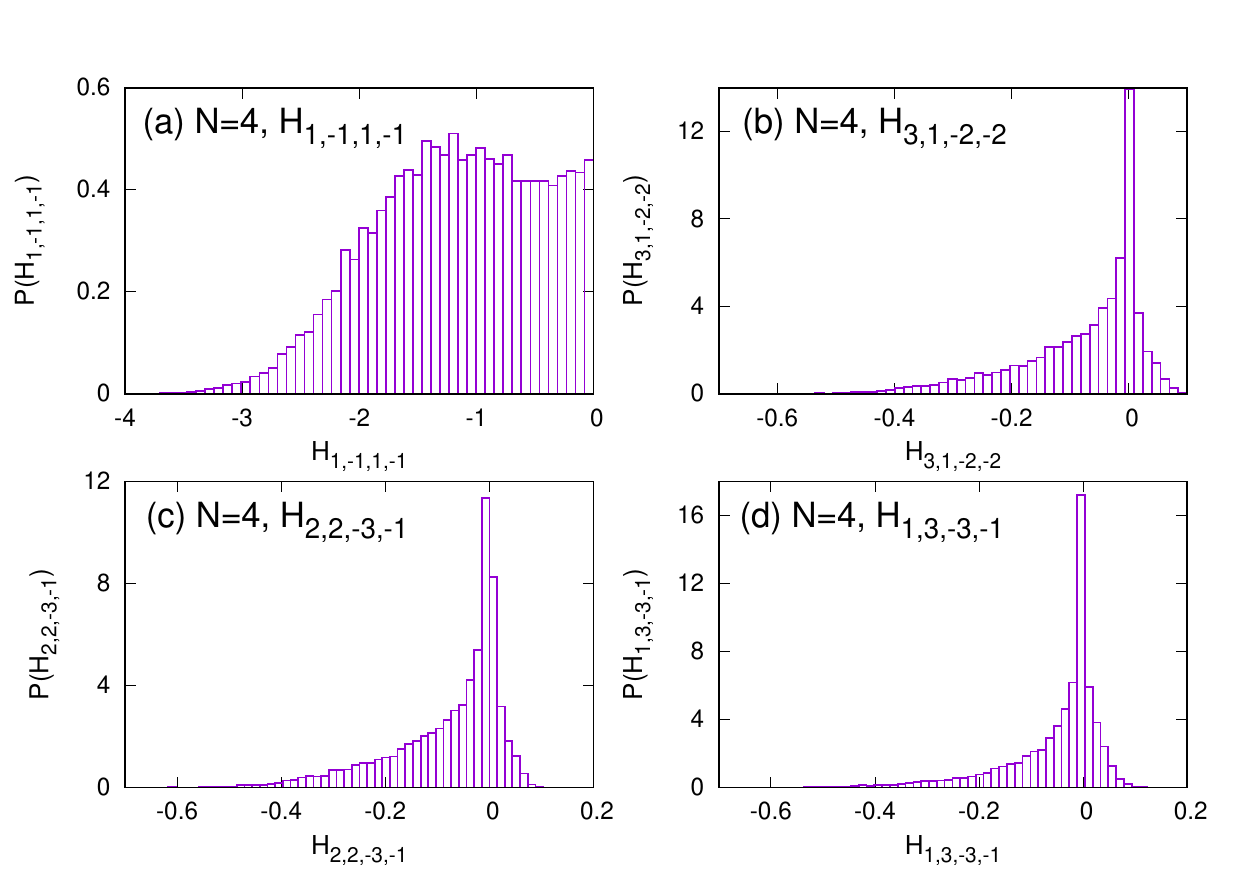}
    \caption{ The figure shows the distribution
      $P(H_{n_1,\,\,n_2,\,\,n_3,\,\,n_4})$ for a quantum dot connected to
      $N=4$ superconducting leads, where $H_{n_1,\,\,n_2,\,\,n_3,\,\,n_4}$
      is given by Eq.~(\ref{eq:harmo-n}) with $N=p=4$. Panels a, b, c
      and d correspond to $(n_1,\,\,n_2,\,\,n_3,\,n_4)=(1,\,-1,\,1,\,-1)$,
      $(3,\,1,\,-2,\,-2)$, $(2,\,3,\,-3,\,-1)$ and $(1,\,3,\,-3,\,-1)$
      respectively. The coupling parameters $\Gamma$s entering
      Eqs.~(\ref{eq:H-infinite})-(\ref{eq:E-ABS-gap-infini}) are
      chosen at random in a Gaussian distribution centered at
      $\overline{\Gamma}=1$ and with root-mean-square $\delta \Gamma=0.5$.      
    \label{fig:figure-histos-N4}}
\end{figure*}

\subsection{$p$-terminal dc-Josephson current with $N$ superconducting
  leads}

\label{sec:p-ter}

In this subsection, we provide expression for the $p$-terminal
dc-Josephson current in a device containing $N$ superconducting leads,
with $p\le N$.

We consider the $p$ integers $a_1,\,....,\,\,a_p$ such that $1\le
a_1<a_2<...<a_p\le N$. They encode supercurrent flowing through the
superconducting leads $S_{a_1},\,S_{a_2},\,...,\,S_{a_p}$, and absence
of supercurrent through $S_{b_1},\,S_{b_2},\,...,\,S_{b_{N-p}}$, where
$a_\alpha\ne b_\beta$ for all $\alpha=1,\,2,\,...,\,p$ and
$\beta=1,\,2,\,...,\,\,n-p$.  Then, $m_{a_{\alpha}}\ne 0$ if
$a_{\alpha}$ belong to $\{a_1,\,...,\,a_p\}$, and $m_{b_\beta}= 0$ if
$b_\beta$ is not in $\{a_1,\,...,\,a_p\}$, see
Eqs.~(\ref{eq:IS-N0-a1})-(\ref{eq:IS-N0-a2}).

These statements are illustrated on figure~\ref{fig:cartoon_for_m} for
the $N=4$ superconducting leads $S_1$, $S_2$, $S_3$ and $S_4$.
Figure~\ref{fig:cartoon_for_m}-a1, \ref{fig:cartoon_for_m}-b1 and
\ref{fig:cartoon_for_m}-c1 show the balance in the number of
``outgoing'' and ``incoming'' Cooper pairs, for the
$p=2,\,3,\,4$-terminal dc-Josephson currents respectively, in a device
containing $N=4$ superconducting
leads. Figures~\ref{fig:cartoon_for_m}-a2, \ref{fig:cartoon_for_m}-b2
and \ref{fig:cartoon_for_m}-c2 show the values of the variables
$a_\alpha$ and $m_{a_\alpha}$ for the processes on
figures~\ref{fig:cartoon_for_m}-a1, \ref{fig:cartoon_for_m}-b1 and
\ref{fig:cartoon_for_m}-c1 with $(m_{a_1},\,m_{a_2})=(1,-1)$,
$(m_{a_1},\,m_{a_2},\,m_{a_3})=(3,\,-1,\,-2)$ and
$(m_{a_1},\,m_{a_2},\,m_{a_3},\,m_{a_4})=(1,\,-1,\,1,\,-1)$
respectively.  We also note that dressing the processes on
figures~\ref{fig:cartoon_for_m}-a1 and ~\ref{fig:cartoon_for_m}-a2
according to figures~\ref{fig:cartoon_for_dressing}-a1 and
~\ref{fig:cartoon_for_dressing}-a2 leaves unchanged the values of
$a_\alpha$ and $m_{a_\alpha}$.

Now, Eqs.~(\ref{eq:IS-N0-a1})-(\ref{eq:IS-N0-a2}) are decomposed as
the following:
\begin{equation}
  \label{eq:IS-a}
  I_{N_0}^{S,\,\left(N\right)}\left(\varphi_{1},\,...,\,\varphi_{N}\right)= \sum_{p=2}^N
  I_{N_0}^{S,\left(p\right)}\left(\varphi_{1},\,...,\,\varphi_{N}\right)
  ,
\end{equation}
where
\begin{widetext}
\begin{eqnarray}
  \label{eq:IS-a-bis}
  &&I_{N_0}^{S,\,\left(N\right)\,\left(p\right)}
  \left(\varphi_{1},\,...,\,\varphi_{N}\right)= \sum_{a_1=1}^N
  \sum_{a_2=a_1+1}^N ... \sum_{a_p=a_{p-1}+1}^N
  \delta\left(\prod_{\alpha=1}^N
  \left(1-\delta\left(a_\alpha,\,n_0\right)\right)\right)
  \sum_{m_{a_1}\ne 0}\sum_{m_{a_2}\ne 0} ... \sum_{m_{a_p}\ne 0}
  \times\\ &&
  I_{c,\,n_0}^{\left(N\right)\,\left(p\right)\,\left(a_1,\,a_2,\,...,\,a_p\right)}
  \left[m_{a_1},\,...,\,m_{a_p}\right]
    \sin\left(m_{a_1} \varphi_{a_1} + m_{a_2}\varphi_{a_2} + .... +
    m_{a_p}\varphi_{a_p}\right)
    \delta\left(m_{a_1}+m_{a_2}+...+m_{a_p}\right)
\nonumber
  .
\end{eqnarray}
\end{widetext}
The $\delta$-function $\delta\left(\prod_{\alpha=1}^N
\left(1-\delta_{a_\alpha,\,n_0}\right)\right)$ in
Eq.~(\ref{eq:IS-a-bis}) is used to indicate that the lead $S_{N_0}$
(through which the supercurrent is evaluated) is among
$S_{a_1},\,...,\,S_{a_p}$ (which sustain the considered $p$-terminal
dc-Josephson process). Namely, $a_{\alpha_0}= N_0$ for one of the
$\alpha=\alpha_0$ implies $\prod_{\alpha=1}^N
\left(1-\delta_{a_\alpha,\,n_0}\right)=0$ and $a_\alpha\ne N_0$ for all
$\alpha=1,\,...,p$ implies $\prod_{\alpha=1}^N
\left(1-\delta_{a_\alpha,\,n_0}\right)=1$.

Eqs.~(\ref{eq:IS-N0-a1})-(\ref{eq:IS-N0-a2}) were calculated from the
contribution of the ABS to the dc-Josephson supercurrent. The present
paper is based on the general current-phase relation given by
Eq.~(\ref{eq:IS-a-bis}), which also captures the contribution of the
continua \cite{Pillet1,Pillet2,Benzoni}.

\subsection{Single-level quantum dot}
\label{sec:dot}

In this subsection, we illustrate
Eqs.~(\ref{eq:IS-N0-a1})-(\ref{eq:IS-N0-a2}) for a quantum dot
connected to three or four superconducting leads.  We denote by
$\Gamma_1,\,\Gamma_2,\,...,\,\Gamma_N$ the line-width broadening
associated to hopping between the dot and each of the superconducting
leads $S_1,\,S_2,\,...,\,S_N$ in the normal-state.  We take the
infinite-gap limit, see for instance
Refs.~\onlinecite{topo1-plus-Floquet,Zazunov,Meng}. The $2\times 2$
Nambu Hamiltonian takes the form
\begin{equation}
  \label{eq:H-infinite}
  {\cal H}_\infty=\left(
  \begin{array}{cc} 0 & \sum_{\alpha=1}^N \Gamma_\alpha \exp\left( i \varphi_\alpha
    \right)\\
    \sum_{\alpha=1}^N \Gamma_\alpha \exp\left(-i \varphi_\alpha
    \right) & 0 \end{array}\right)
  .
\end{equation}
The ABS energies
\begin{eqnarray}
  \label{eq:E-ABS-gap-infini}
&&  E_{ABS,\pm}\left(\varphi_1,\,\varphi_2,\,...,\,\varphi_N
  \right)\\
  &=&\pm \left|\sum_{\alpha=1}^N \Gamma_\alpha \exp\left( i
  \varphi_\alpha \right)\right| .
  \nonumber
\end{eqnarray}
imply nonvanishingly small Fourier coefficients
$E^{\{m\}}_{ABS,\pm}\left[m_1,\,m_2,\,...,\,m_N\right]$ in
Eq.~(\ref{eq:E-ABS1}). This yields the $p$-terminal Josephson effect,
{\it i.e.}  components of the supercurrent which are sensitive to the
superconducting phase variables of $p$ superconducting leads:
\begin{equation}
  \label{eq:Psi-p}
  \Psi_p\left[a_1,\,a_2,\,...,\,a_p\right]
=  m_{a_1}\varphi_{a_1}+m_{a_2}\varphi_{a_2}+
  ...+m_{a_p}\varphi_{a_p}
  ,
\end{equation}
where $\{a_\alpha\}$ is a set of $p$ integers such that $1\le
a_1<a_2<...<a_p\le N$ and $m_{a_\alpha}\ne 0$ for all
$\alpha=1,\,...,\,p$.

Figures~\ref{fig:figure-histos-N3} and~\ref{fig:figure-histos-N4} show
the distributions $P(H_{n_1,\,n_2,\,n_3})$ and $P(H_{n_1,\,n_2,\,n_3,\,n_4})$ for
the Fourier harmonics of the three and four-terminal dc-Josephson
currents
\begin{eqnarray}
  \label{eq:harmo-n}
  H_{n_{a_1},\,\,n_{a_2},\,...,\,n_{a_p}}&=&
  \int \frac{d\varphi_{a_1}}{2\pi} 
  \int \frac{d\varphi_{a_2}}{2\pi}
  ...
  \int \frac{d\varphi_{a_p}}{2\pi}\times\\
&&  I_1^{S(N)(p)}\left(
  \varphi_{a_1},\,\varphi_{a_2},\,...,\varphi_{a_p}\right)\times \\
  &&\sin\left(m_{a_1} \varphi_{a_1}
  + m_{a_2} \varphi_{a_2} + ... +
  m_{a_p} \varphi_{a_p}\right)
  \nonumber
  ,
  \nonumber
\end{eqnarray}
where $a_\alpha\equiv \alpha$ on the examples of
figures~\ref{fig:figure-histos-N3} and~\ref{fig:figure-histos-N4}
where $p=N$.

The distribution of the $\Gamma$s in
Eqs.~(\ref{eq:H-infinite})-(\ref{eq:E-ABS-gap-infini}) is
{taken as a} Gaussian centered at
$\overline{\Gamma}=1$, with root-mean-square $\delta \Gamma=0.5$. The
negative values of the $\Gamma$s are rejected in the numerical
calculation. These parameter values imply that $H_{n_1,\,n_2,\,n_3}$
and $H_{n_1,\,n_2,\,n_3,\,n_4}$ are roughly of order unity on
figures~\ref{fig:figure-histos-N3}
and~\ref{fig:figure-histos-N4}. Figure~\ref{fig:figure-histos-N3}a
shows the $(1,\,0,\,-1)$ harmonics with $N=3$, with the expected
positive sign.  Figure~\ref{fig:figure-histos-N3}b shows the expected
negative sign of the $(2,\,0,\,-2)$
harmonics. Figure~\ref{fig:figure-histos-N3}c shows the $(1,\,1,\,-2)$
quartets with negative sign corresponding to the negative
$I_{c,\,dCAR}<0$ mentioned in
section~\ref{sec:connection}. Figure~\ref{fig:figure-histos-N3}d shows
the $(2,\,1,\,-3)$
harmonics. Figures~\ref{fig:figure-histos-N4}a-\ref{fig:figure-histos-N4}d
further illustrate the multiterminal dc-Josephson effect on the
example of $N=4$ superconducting leads, with
$(n_1,\,\,n_2,\,\,n_3,\,n_4)=(1,\,-1,\,1,\,-1)$, $(3,\,1,\,-2,\,-2)$,
$(2,\,3,\,-3,\,-1)$ and $(1,\,3,\,-3,\,-1)$ respectively.

It is concluded that the multiterminal dc-Josephson effect naturally
emerges if a quantum dot is connected to $N$ superconducting leads, in
the form of the higher-order harmonics of the dc-Josephson
current-phase relation, see the selected values of $(n_1,\,\,n_2,\,\,n_3)$
and $(n_1,\,\,n_2,\,\,n_3,\,\,n_4)$ for $N=3$ and $N=4$ superconducting
leads on figures~\ref{fig:figure-histos-N3}
and~\ref{fig:figure-histos-N4} respectively.

\section{Examples with $N=3,\,4$ superconducting leads}
\label{sec:N=34examples}
In this section, we present the specific examples of $N=3$ and $N=4$
superconducting leads, see the devices on
figures~\ref{fig:N-terminal-junction}b
and~\ref{fig:N-terminal-junction}c respectively. The cases $N=3$ and
$N=4$ are treated in sections~\ref{sec:N3-example}
and~\ref{sec:N4-example} respectively. We show that $\chi^{(3)}$ for
$N=3$ yields the $p=2$ and the $p=3$-terminal dc-Josephson effect, and
that $\chi^{(4)}$ for $N=4$ yields the $p=3$ and the $p=4$-terminal
dc-Josephson effect, whatever the value of the $p=2$-terminal
dc-Josephson current. Thus, our proposal has the potential for
directly probing the $p=3$ or $p=4$-terminal dc-Josephson effect with
$N=4$ superconducting leads in microwave experiments.

\subsection{Device with $N=3$ superconducting leads on
  figure~\ref{fig:N-terminal-junction}b}
\label{sec:N3-example}

In this subsection, we consider the device with $N=3$ superconducting
leads on figure~\ref{fig:N-terminal-junction}b. We specifically
evaluate the flux-$\Phi$ sensitivity of the supercurrent through the lead
$S_1$:
\begin{eqnarray}
\nonumber
&&  I_1^{(3)}\left(\varphi_1,\varphi_2,\varphi_3\right)\\&=&
  \label{eq:I-panelb-debut-1-ori}
  \sum_{n \ne 0} I_c^{'\left(3\right)\,\left(2\right)\,\left(1,\,2\right)}[n]
  \sin\left[ n \left(\varphi_1-\varphi_2 \right)\right]\\ &+&
   \label{eq:I-panelb-debut-1-ori-2}
 \sum_{n \ne 0} I_c^{'\left(3\right)\,\left(2\right)\,\left(1,\,3\right)}[n]
  \sin\left[ n \left(\varphi_1-\varphi_3 \right)\right] \\
  \label{eq:I-panelb-debut-1-ori-fin}
  &+& \sum_{n_1\ne 0,\,\,n_2\ne 0}
  I_c^{'\left(3\right)\,\left(3\right)\,\left(1,\,2,\,3\right)}[n_1,\,\,n_2]
  \times\\ && \sin\left[ n_1 \left(\varphi_1-\varphi_3\right) +n_2
    \left(\varphi_2-\varphi_3\right)\right] .\nonumber
  \label{eq:I-panelb-debut-1-ori-fin-2}
\end{eqnarray}
Taking into account that $\varphi_3=\varphi_2-\Phi$, we obtain
$\varphi_1-\varphi_3= \varphi_1-\varphi_2+\Phi$.  The considered
signal for $N=3$ is defined as the following:
\begin{eqnarray}
    \label{eq:S-(3)}
{\chi}^{(3)}(\varphi_1-\varphi_2,\Phi) =\frac{\partial
  I_1^{(3)}(\varphi_1-\varphi_3,\varphi_2-\varphi_3)}{\partial
  \left(\varphi_2-\varphi_3\right)}
,
\end{eqnarray}
which is analogous to $\chi^{(4)}$ in the above
Eqs.~(\ref{eq:S-(4)-gene})-(\ref{eq:S-(4)-gene-bis}) with $N=4$
superconducting leads. {Again, $\chi^{(3)}$ in
  Eq.~(\ref{eq:S-(3)}) can be detected in microwave experiments.}

The following is deduced from
Eqs.~(\ref{eq:I-panelb-debut-1-ori})-(\ref{eq:I-panelb-debut-1-ori-fin-2}):
\begin{eqnarray}
  \nonumber
  && {\chi}^{(3)}(\varphi_1-\varphi_2,\Phi)\\
  \label{eq:partial1}
  &=& \sum_{n \ne 0} n
    I_c^{'\left(3\right)\,\left(2\right)\,\left(1,\,3\right)}[n]
    \cos\left[ n \left(\varphi_1-\varphi_2+\Phi
      \right)\right]\\ \label{eq:partial1-2d} &+& \sum_{n_1,\,n_2} n_2
    I_c^{'\left(3\right)\,\left(3\right)\,\left(1,\,2,\,3\right)}[n_1,\,n_2]\times\\ &&
    \cos\left[ n_1 \left(\varphi_1-\varphi_2+\Phi\right) +n_2
      \Phi\right] .\nonumber
\end{eqnarray}
The zero-flux limit $\Phi=0$ of
Eqs.~(\ref{eq:partial1})-(\ref{eq:partial1-2d}) is the following:
\begin{eqnarray}
    \label{eq:partial1-0}
    && {\chi}^{(3)(0)}(\varphi_1-\varphi_2) \equiv
          {\chi}^{(3)}(\varphi_1-\varphi_2,0) \\ \nonumber &=& \sum_{n
            \ne 0} n \left[
            I_c^{'\left(3\right)\,\left(2\right)\,\left(1,\,3\right)}[n]
            +\sum_{n_2\ne 0}
            I_c^{'\left(3\right)\,\left(3\right)\,\left(1,\,2,\,3\right)}[n,\,n_2]
            \right] \times\\ &&\cos\left[ n \left(\varphi_1-\varphi_2
            \right)\right] . \label{eq:partial1-0-fin}
\end{eqnarray}
Both $p=2$ and $p=3$ contribute to ${\chi}^{(3)}$ in
Eqs.~(\ref{eq:partial1})-(\ref{eq:partial1-2d}) and ${\chi}^{(3)(0)}$
in Eqs.~(\ref{eq:partial1-0})-(\ref{eq:partial1-0-fin}). Thus,
${\chi}^{(3)}$ or ${\chi}^{(3)(0)}$ with $N=3$ superconducting leads
cannot be used to probe the $p=3$-terminal dc-Josephson effect
independently on the contribution of the $p=2$-terminal Josephson
effect, see the discussion in section~\ref{sec:connection}.

\subsection{Device with $N=4$ superconducting leads on
  figure~\ref{fig:N-terminal-junction}c}
\label{sec:N4-example}

In this subsection, we now consider the device with $N=4$
superconducting leads, see figure~\ref{fig:N-terminal-junction}c. We
evaluate the sensitivity of the supercurrent $I_1^{(4)}$ through the
lead $S_1$ on the three superconducting phase differences
$\varphi_1-\varphi_4$, $\varphi_2-\varphi_4$ and
$\varphi_3-\varphi_4$.
\begin{widetext}
The supercurrent through the superconducting lead $S_1$ takes the form
\begin{eqnarray}
  \label{eq:I-panelb-debut-1}
  I_1^{(4)}&=& \sum_{n \ne 0} I_c^{'\left(4\right)\,\left(2\right)\,\left(1,\,2\right)}[n] \sin\left[ n
    \left(\varphi_1-\varphi_2 \right)\right]+ \sum_{n \ne 0}
  I_c^{'\left(4\right)\,\left(2\right)\,\left(1,\,3\right)}[n] \sin\left[ n
    \left(\varphi_1-\varphi_3 \right)\right] + \sum_{n \ne 0}
  I_c^{'\left(4\right)\,\left(2\right)\,\left(1,\,4\right)}[n] \sin\left[ n
    \left(\varphi_1-\varphi_4 \right)\right]
  \label{eq:BB1}\\ &&+
  \label{eq:box-page-6}
  \sum_{n_1\ne 0,\,\,n_2\ne 0}
  I_c^{'\left(4\right)\,\left(3\right)\,\left(1,\,2,\,3\right)}[n_1,\,\,n_2]
  \sin\left[ n_1 \left(\varphi_1-\varphi_3\right) +n_2
    \left(\varphi_2-\varphi_3\right)\right] \\ &&+
  \sum_{n_1\ne 0,\,\,n_2\ne 0}
  I_c^{'\left(4\right)\,\left(3\right)\,\left(1,\,2,\,4\right)}[n_1,\,\,n_2]
  \sin\left[ n_1 \left(\varphi_1-\varphi_4\right) +n_2
    \left(\varphi_2-\varphi_4\right)\right] \\
\label{eq:AA2}
  &&+ \sum_{n_1\ne 0,\,\,n_3\ne 0}
I_c^{'\left(4\right)\,\left(3\right)\,\left(1,\,3,\,4\right)}[n_1,\,\,n_3]
\sin\left[ n_1 \left(\varphi_1-\varphi_4\right) +n_3
  \left(\varphi_3-\varphi_4\right)\right]
\\ &&+\label{eq:star-page-7} \sum_{n_1\ne 0,\,\,n_2\ne 0,\,\,n_3\ne 0}
I_c^{'\left(4\right)\,\left(4\right)\,\left(1,\,2,\,3,\,4\right)}[n_1,\,\,n_2,\,\,n_3]
\sin\left[ n_1 \left(\varphi_1-\varphi_4\right) + n_2
  \left(\varphi_2-\varphi_4\right) +n_3
  \left(\varphi_3-\varphi_4\right)\right]
.
\end{eqnarray}

The fluxes $\Phi_A$ and $\Phi_B$ are taken into account with
$\varphi_3=\varphi_2-\Phi_A$ and $\varphi_4=\varphi_3-\Phi_B$, which
leads to
\begin{eqnarray}
\label{eq:expression1}
  \varphi_3-\varphi_4&=&\Phi_B\\
  \varphi_2-\varphi_4&=&\Phi_A+\Phi_B
  .
  \label{eq:expression2}
\end{eqnarray}

Eq.~(\ref{eq:box-page-6}) and Eq.~(\ref{eq:star-page-7}) contribute
for finite values to $\chi^{(4)}\ne 0$ in
Eqs.~(\ref{eq:S-(4)-gene})-(\ref{eq:S-(4)-gene-bis}) and
$\chi^{(4)(0)}\left(\varphi_1-\varphi_4\right) \ne 0$ in
Eqs.~(\ref{eq:S-(4)})-(\ref{eq:S-(4)-bis}). The other terms in
Eqs.~(\ref{eq:I-panelb-debut-1})-(\ref{eq:star-page-7}) do not
contribute to $\chi^{(4)}$ and
$\chi^{(4)(0)}\left(\varphi_1-\varphi_N\right)$ after the partial
derivatives with respect to $\varphi_2-\varphi_4$ and
$\varphi_3-\varphi_4$ have been taken:
\begin{eqnarray}
  \label{eq:TOTO1}
  {\chi}^{(4)}(\varphi_1-\varphi_4,\Phi_A,\Phi_B) &=&
  \sum_{n_1\ne 0,\,\,n_2\ne 0} n_2\left(n_1+n_2\right)
  I_c^{'\left(4\right)\,\left(3\right)\,\left(1,\,2,\,3\right)}[n_1,\,\,n_2]\times\\
  \nonumber&& \sin\left[ n_1
    \left(\varphi_1-\varphi_4\right) +n_2 \left(\Phi_A+\Phi_B\right)-
    \left(n_1+n_2\right) \Phi_B \right] \\ &-& \sum_{n_1\ne 0,\,\,n_2\ne 0,\,\,n_3\ne 0}
  n_2 n_3
  I_c^{'\left(4\right)\,\left(4\right)\,\left(1,\,2,\,3,\,4\right)}[n_1,\,\,n_2,\,\,n_3]
  \sin\left[ n_1 \left(\varphi_1-\varphi_4\right) + n_2
    \left(\Phi_A+\Phi_B\right) +n_3 \Phi_B\right]  ,
  \label{eq:TOTO1-fin}
\end{eqnarray}
where we used Eqs.~(\ref{eq:I-panelb-debut-1})-(\ref{eq:expression2}).

Assuming zero flux $\Phi_A=\Phi_B=0$ (or, equivalently,
$\varphi_2-\varphi_4=0$ and $\varphi_3-\varphi_4=0$),
Eqs.~(\ref{eq:TOTO1})-(\ref{eq:TOTO1-fin}) simplify as
\begin{eqnarray}
  \label{eq:df1}
{\chi}^{(4)(0)}(\varphi_1-\varphi_4) &\equiv&
        {\chi}^{(4)}(\varphi_1-\varphi_4,0,0)\\ &=& \sum_{n_1\ne
          0,\,\,n_2\ne 0} n_2\left[\left(n_1+n_2\right)
          I_c^{'\left(4\right)\,\left(3\right)\,\left(1,\,2,\,3\right)}[n_1,\,\,n_2]
          -\sum_{n_3\ne 0} n_3
          I_c^{'\left(4\right)\,\left(4\right)\,\left(1,\,2,\,3,\,4\right)}[n_1,\,\,n_2,\,\,n_3]
          \right] \sin\left[ n_1 \left(\varphi_1-\varphi_4\right)
          \right] .\label{eq:df2}
  \label{eq:TOTO1-0}
\end{eqnarray}
Eqs.~(\ref{eq:TOTO1})-(\ref{eq:TOTO1-0}) contain the $p=3$ and the
$p=4$-terminal contributions
$I_c^{'\left(4\right)\,\left(3\right)\,\left(1,\,2,\,3\right)}$ and
$I_c^{'\left(4\right)\,\left(4\right)\,\left(1,\,2,\,3,\,4\right)}$. The
$p=2$ two-terminal
$I_c^{'\left(4\right)\,\left(2\right)\,\left(1,\,2\right)}$,
$I_c^{'\left(4\right)\,\left(2\right)\,\left(1,\,3\right)}$,
$I_c^{'\left(4\right)\,\left(2\right)\,\left(1,\,4\right)}$ do not
appear in Eqs.~(\ref{eq:TOTO1})-(\ref{eq:TOTO1-0}). Thus, experimental
detection of $\chi^{(4)}\ne 0$ or
$\chi^{(4)}(0)\left(\varphi_1-\varphi_4\right)\ne 0$ implies evidence
for the $p=3$ or $p=4$ dc-Josephson effects. Specifically, we
demonstrated the following for $N=4$ superconducting leads (see
figure~\ref{fig:N-terminal-junction}c):

``{\it Nonvanishingly small ${\chi}^{(4)}
  \left(\varphi_1-\varphi_4,\Phi_A,\Phi_B\right)\ne 0$ or ${\chi}^{(4)(0)}\left(\varphi_1-\varphi_4\right)\ne 0$ with $N=4$
  superconducting leads}''

generically implies

``{\it Evidence for the $p=3$ or $p=4$-terminal Josephson effect}''.

Now, we generalize this statement to a device containing an arbitrary
number $N$ of the superconducting leads, see
figures~\ref{fig:N-terminal-junction}b, \ref{fig:N-terminal-junction}c
and \ref{fig:N-terminal-junction}d for $N=3,\,4,\,5$ respectively, and
figure~\ref{fig:multiloop} for $N=6$.

\section{$N$ superconducting leads}
\label{sec:general-N}

We demonstrated in the previous section~\ref{sec:N=34examples} that
$\chi^{(3)}\ne 0$ or $\chi^{(3)(0)}(\varphi_1-\varphi_3)\ne 0$ for
$N=3$, and that $\chi^{(4)} \ne 0$ or
$\chi^{(4)(0)}(\varphi_1-\varphi_4)\ne 0$ for $N=4$ imply evidence for
the $p=2,\,3$ and the $p=3,\,4$ dc-Josephson effects respectively.

Now, we generalize this statement to arbitrary values of the number
$N$ of the superconducting leads. Subsection~\ref{sec:arbitrary-N}
provides expression of the higher-order nonlocal inverse inductance
$\chi^{(N)}$ at arbitrary $N$. Subsections~\ref{sec:even}
and~\ref{sec:odd} present the calculations for even and odd values of
$N$ respectively.

  \subsection{Higher-order
    nonlocal inverse inductance}
\label{sec:arbitrary-N}

In this subsection, we discuss how $N=3,\,4$ in the previous
sections~\ref{sec:N3-example} and~\ref{sec:N4-example} respectively
can be generalized to arbitrary number $N$ of superconducting
leads. 

Eqs.~(\ref{eq:IS-a})-(\ref{eq:IS-a-bis}) for the dc-Josephson current
are written as
\begin{eqnarray}
  \label{eq:IprimeS-app-1}
&&
  I^{'S\,\left(N\right)}_1\left(\varphi_{1}-\varphi_{N},
  \,\varphi_{2}-\varphi_{N},\,...,
  \varphi_{N-1}-\varphi_{N}\right) =\sum_{p=2}^{N-1}
  \sum_{a'_1=1}^{N-1} \sum_{a'_2=a'_1+1}^{N-1}
  ... \sum_{a'_p=a'_{p-1}+1}^{N-1}
  \\ &&\delta\left(\prod_{\alpha=1}^{N-1}\left(1 -
  \delta\left(a'_\alpha,1\right)\right)\right)\sum_{m'_{a'_1}\ne 0}
  \sum_{m'_{a'_2}\ne 0} ... \sum_{m'_{a'_p}\ne 0}
  I_c^{'\left(N\right)\,\left(1\right)\,\left(p\right)\,\left(a'_1,\,a'_2,\,...,\,a'_p\right)}[m'_{a'_1},\,...,\,m'_{a'_p}]
  \sin\left(\sum_{\alpha=1}^p m'_{a'_\alpha}
  \left(\varphi_{a'_\alpha}-\varphi_N\right)\right)
  \nonumber
,
\end{eqnarray}
\end{widetext}
where we evaluated the supercurrent through the lead labeled by
$N_0=1$, and used the phase $\varphi_N$ of the superconducting lead
$S_N$ as a reference. The variable $a'_\alpha$ in
Eq.~(\ref{eq:IprimeS-app-1}) is defined in the interval $1\le
a'_\alpha \le N-1$ while $a_\alpha$ in Eq.~(\ref{eq:IS-a-bis}) is such
that $1\le a_\alpha \le N$.

We define the order-$\left(N-2\right)$ nonlocal inductance as the
signal which is proposed to be detected in microwave experiments:
\begin{eqnarray}
\nonumber&&  {\chi}^{\left(N\right)} \equiv
 \left[L^{-1}_{N-2}\right]'\left(\varphi_{1}-\varphi_{N},\,
\varphi_{2}-\varphi_{N},\,..., \varphi_{N-1}-\varphi_{N}\right)\\&=&
\frac{\partial^{N-2}}{ \partial\left(\varphi_2-\varphi_N\right)
  \partial\left(\varphi_3-\varphi_N\right) ...
  \partial\left(\varphi_{N-1}-\varphi_N\right)}\times\nonumber
\\
&&I^{'S}_1\left(\varphi_{1}-\varphi_{N},\,\varphi_{2}-\varphi_{N},\,...,
\varphi_{N-1}-\varphi_{N}\right)
  \label{eq:A1}
.
\end{eqnarray}

We note that the lead $S_N$ is singled out in Eq.~(\ref{eq:A1}): the
superconducting phase variable $\varphi_N$ is the reference for
evaluating the partial derivatives of $I^{'S}_1$ with respect to all
of the $\varphi_n-\varphi_N$, with $n=2,\,...,\,n-1$. Now, we provide a
calculation of Eq.~(\ref{eq:A1}).

\begin{widetext}

\begin{figure*}[htb]
  \begin{minipage}{.7\textwidth}
  \includegraphics[width=.8\textwidth]{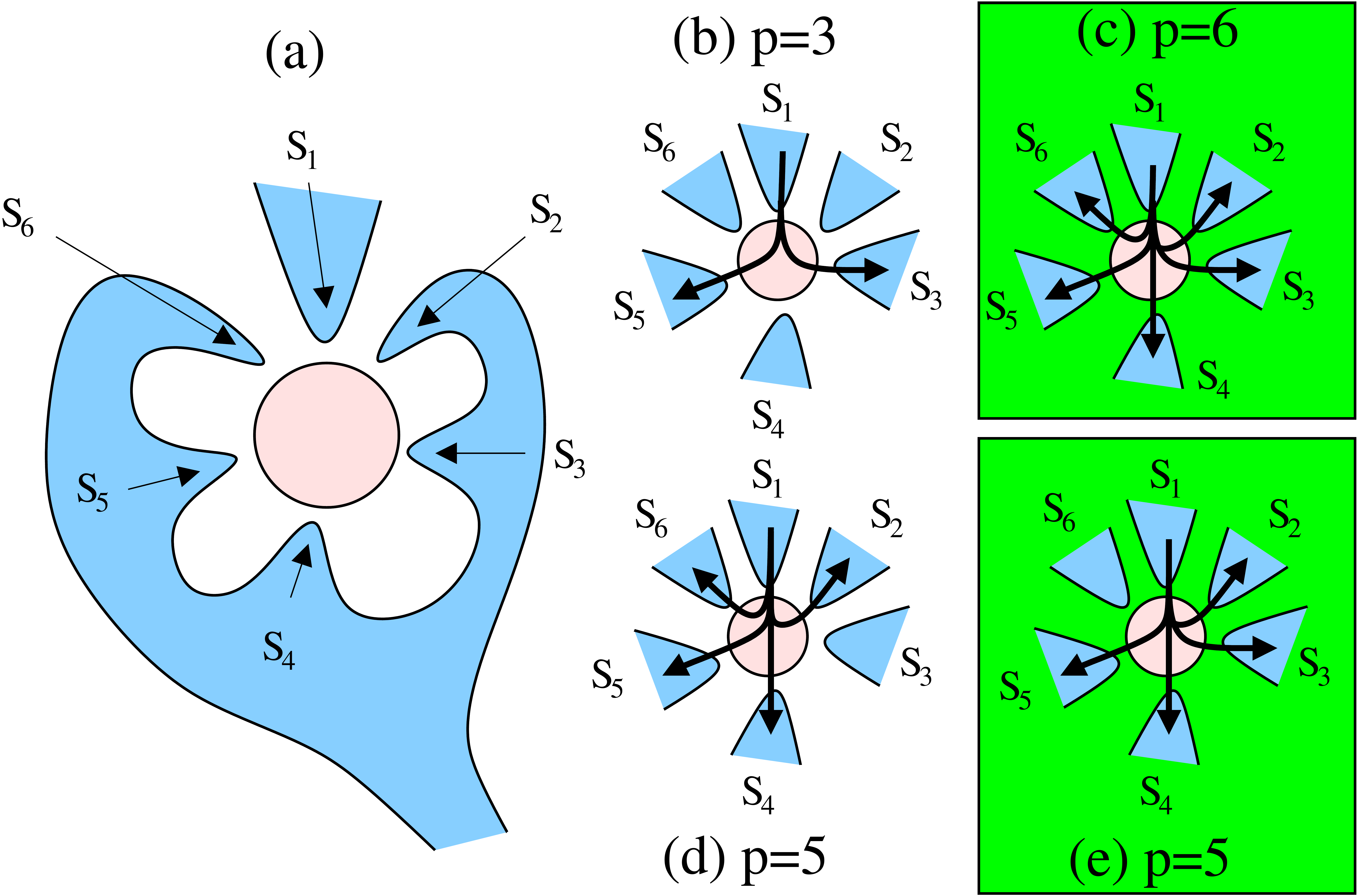}\end{minipage}\begin{minipage}{.28\textwidth}
  \caption{The figure shows a $N=6$ dc-Josephson junction (a). Panels
    b, c and d, e show the $p=3,\,6,\,5,\,5$-terminal dc-Josephson
    processes in a device containing $N=6$ superconducting leads
    respectively. The processes on the green-highlighted panels c and
    e contribute for a nonvanishingly small
    $\left[\tilde{L}^{-1}_{N-2}\right]^{'\left(N\right)} \ne 0$ in
    Eq.~(\ref{eq:A1}), on the condition of nonvanishingly small
    corresponding critical currents.
    \label{fig:multiloop}}
  \end{minipage}
\end{figure*}

\subsection{Even number $N=2q$ of superconducting leads}
\label{sec:even}

In this subsection, we assume that $N=2q$ is even. In addition, we
assume in subsection~\ref{sec:ne0} that $\sum_{\alpha=1}^{2q-1}
m'_{b'_\alpha}\ne 0$ (with $m'_{b'_\beta}\ne 0$ for all
$\beta=1,\,...,2q-1$). Conversely, subsection~\ref{sec:eq0} deals with
$\sum_{\alpha=1}^{2q-1} m'_{b'_\alpha}= 0$.

\subsubsection{$p=2q$-terminal Josephson processes with $N=2q$
  superconducting leads}
\label{sec:ne0}

In this subsection, we assume that $N=2q$ is even and
$\sum_{\alpha=1}^{2q-1}
m'_{b'_\alpha}\ne 0$ (with $m'_{b'_\beta}\ne 0$ for all
$\beta=1,\,...,2q-1$).

This results in the $p=2q$-terminal dc-Josephson contribution
$\left[L^{-1}_{2q-2}\right]^{'(2q)}$ to ${\chi}^{(2q)}=
\left[L^{-1}_{2q-2}\right]'$ in Eq.~(\ref{eq:A1}). Taking into account
that $b'_\alpha=\alpha$ for a $p=2q=N$-terminal dc-Josephson process,
we find
\begin{eqnarray}
  \label{eq:IprimeS-2-even-A-0}
  \left[L^{-1}_{2q-2}\right]^{'(2q)}&=&\left(-\right)^{q+1}
  \sum_{m'_1\ne 0} \sum_{m'_2\ne 0} ... \sum_{m'_{2q-1}\ne 0}
  I_{c,1}^{'\left(2q\right)\,\left({2q}\right)\,\left(1,2,\,...,{{2q-1}}\right)}
  [m'_{1},\,...,\,m'_{{{2q}-1}}]\times\\ &&
  \left(\prod_{\alpha=2}^{{2q}-1} m'_{\alpha}\right)
  \sin\left(\sum_{\alpha=1}^{{2q}-1} m'_{{\alpha}}
  \left(\varphi_{{\alpha}}-\varphi_{{2q}}\right)\right) .  \nonumber
\end{eqnarray}

Specializing to
$\varphi_2-\varphi_{2q}=...=\varphi_{{2q}-1}-\varphi_{2q}=0$ and
$\varphi_1-\varphi_{2q}\ne 0$ leads to
\begin{eqnarray}
  \label{eq:IprimeS-2-even-2-A}
  \left[\tilde{L}^{-1}_{{2q}-2}\right]^{'(2q)(0)} &=&\left(-\right)^{q+1}
  \sum_{m'_{1}\ne 0} \sum_{m'_{2}\ne 0}
  ... \sum_{m'_{{{2q}-1}}\ne 0}
  I_c^{'\left(1\right)\,\left({2q}\right)\,\left(1,2,\,...,{{2q-1}}\right)}
  [m'_{1},\,...,\,m'_{{{2q}-1}}] \times\\
  \nonumber
&&  \left(\prod_{\alpha=2}^{{2q}-1} m'_{\alpha}
  \right)\sin\left(m'_{1}
  \left(\varphi_{1}-\varphi_{{2q}}\right)\right) .
\end{eqnarray}

\subsubsection{$p=\left(2q-1\right)$-terminal Josephson processes
  with $N=2q$ superconducting leads}
\label{sec:eq0}

Now, we assume in this subsection that $N=2q$ is even and
\begin{equation}
  \label{eq:assumption1}
  \sum_{\alpha=1}^{{2q}-1} m'_{c'_{\alpha}}= 0
  ,
\end{equation}
with $c'_\alpha=\alpha$ for all $\alpha=1,\,...,\,2q-1$. Then, we
make use of the identities
\begin{eqnarray}
  \label{eq:totoA1}
  &&
  \sin\left(\sum_{\alpha=1}^{2q-1} m'_{{\alpha}}
  \left(\varphi_{\alpha}-\varphi_{2q}\right)\right)
=\sin\left(\sum_{\alpha=1}^{2q-1} m'_{\alpha}
\varphi_{\alpha}\right) =\sin\left(\sum_{\alpha=2}^{2q-1}
m'_{\alpha} \left(\varphi_{\alpha}-\varphi_1\right)\right)
\\&=&\sin\left(\sum_{\alpha=2}^{2q-1} m'_{\alpha}
\left(\varphi_{\alpha}-\varphi_{2q}\right) -\sum_{\alpha=2}^{{2q}-1} m'_{\alpha}
\left(\varphi_{{2q}}-
\varphi_1\right)\right)=
\sin\left(\sum_{\alpha=2}^{{2q}-1} m'_{\alpha}
\left(\varphi_{\alpha}-\varphi_{2q}\right) +m'_{1} \left(\varphi_{{2q}}-
\varphi_1\right)\right)
\nonumber
.
\end{eqnarray}
Specializing to $\varphi_1-\varphi_{2q}\ne 0$ and
$\varphi_{2}-\varphi_{2q}=...=\varphi_{{{2q}-1}}-\varphi_{2q}=0$, we
deduce the contribution
$\left[\tilde{L}^{-1}_{{2q}-2}\right]^{'(2q-1)}$ of the $p=2q-1$-terminal
dc-Josephson processes to $\left[\tilde{L}^{-1}_{{2q}-2}\right]'$ in
the presence of $N=2q$ superconducting leads:
\begin{eqnarray}
  \label{eq:IprimeS-2-even-2-B}
\left[\tilde{L}^{-1}_{{2q}-2}\right]^{(2q-1)(0)}&=&\left(-\right)^{q}
  \sum_{m'_{2}\ne 0}\sum_{m'_{3}\ne 0} ... \sum_{m'_{{{2q}-2}}\ne 0}
  I_{c,1}^{'\left(2q\right)\,
    \left(1\right)\,\left({2q-1}\right)\,\left(1,2,\,...,{{2q}-1}\right)}
  \left[-\left(\sum_{\alpha=2}^{2q-1}m'_\alpha\right),
    \,m'_2,\,...,\,m'_{{{2q}-1}}\right]\times\\ \nonumber &&
  \left(\prod_{\alpha=2}^{{2q}-1} m'_{\alpha} \right)
  \sin\left(\left(\sum_{\alpha=2}^{2q-1} m'_\alpha\right)
  \left(\varphi_1-\varphi_{2q}\right)\right) ,
\end{eqnarray}
where we used $m'_{{2q-1}}=-\sum_{\alpha=1}^{2q-2} m'_{\alpha}$
deduced from Eq.~(\ref{eq:assumption1}).

\subsubsection{Summary}

The zero-flux limit of the predicted experimental signal with $N=2q$
generalizes the above Eq.~(\ref{eq:TOTO1-0}) with $q=2$:
\begin{equation}
  {\chi}^{(2q)(0)}\left(\varphi_1-\varphi_{2q}\right)
  =\left(\ref{eq:IprimeS-2-even-2-A}\right)
  + \left(\ref{eq:IprimeS-2-even-2-B}\right)
  .
\end{equation}
  
\subsection{Odd number $N=2q'+1$ of superconducting leads}
\label{sec:odd}
In this subsection, we assume now that $N=2q'+1$ is odd. We assume in
subsection~\ref{sec:ne0-bis} that $\sum_{\alpha=1}^{2q-1}
m'_{b'_\alpha}\ne 0$ (with $m'_{b'_\beta}\ne 0$ for all
$\beta=1,\,...,2q-1$) and we assume in subsection~\ref{sec:eq0-bis}
that $\sum_{\alpha=1}^{2q-1} m'_{b'_\alpha}= 0$.

\subsubsection{$p=\left(2q'+1\right)$-terminal Josephson processes
with $N=2q'+1$ superconducting leads}
\label{sec:ne0-bis}

In this subsection, we assume that $N=2q'+1$ is odd and
$\sum_{\alpha=1}^{2q'} m'_{b'_\alpha}\ne 0$ (with $m'_{b'_\beta}\ne 0$
for all $\beta=1,\,...,2q'$).

We find
\begin{eqnarray}
  \label{eq:IprimeS-2-odd}
  \left[L^{-1}_{2q'-1}\right]^{'\left(2q'+1\right)}&=& \left(-\right)^{q'+1}
\sum_{m'_1\ne 0} \sum_{m'_2\ne 0}
... \sum_{m'_{2q'}\ne 0}
I_{c,1}^{'\left(2q'+1\right)\,\left(2q'\right)\,\left(1,2,\,...,2q'\right)}
    [m'_1,\,...,\,m'_{2q'}]
\times\\
&&\left(\prod_{\alpha=2}^{2q'} m'_\alpha\right)
    \cos\left(\sum_{\alpha=1}^{2q'} m'_{\alpha}
    \left(\varphi_\alpha-\varphi_{2q'+1}\right)\right)
    .
\end{eqnarray}

Specializing Eq.~(\ref{eq:IprimeS-2-odd}) to
$\varphi_2-\varphi_{2q'+1}=...=\varphi_{2q'}-\varphi_{2q'+1}=0$ and
$\varphi_1-\varphi_{2q'+1}\ne 0$ leads to
\begin{eqnarray}
  \label{eq:IprimeS-2-odd-2-A}
    \left[\tilde{L}^{-1}_{2q'-1}\right]^{'\left(2q'+1\right)(0)} &=&
    \left(-\right)^{q'+1} \sum_{m'_1\ne 0} \sum_{m'_2\ne 0}
    ... \sum_{m'_{2q'}\ne 0}
    I_{c,1}^{'\left(2q'+1\right)\,\left(2q'\right)\,\left(1,2,\,...,2q' \right)}
    [m'_1,\,...,\,m'_{2q'}]\times\\&&\nonumber\left(
    \prod_{\alpha=2}^{2q'} m'_\alpha \right) \cos\left(m'_{1}
    \left(\varphi_1-\varphi_{2q'+1}\right) \right) .
\end{eqnarray}

\subsubsection{$p=2q'$-terminal Josephson processes
  with $N=2q'+1$ superconducting leads}
\label{sec:eq0-bis}

In this subsection, we assume that $N=2q'+1$ is odd and
$\sum_{\alpha=1}^{2q'} m'_{b'_\alpha}= 0$ (with $m'_{b'_\beta}\ne 0$
for all $\beta=1,\,...,2q'$).

We specialize to
$\varphi_2-\varphi_{2q'+1}=...=\varphi_{{2q'}}-\varphi_{2q'+1}=0$ and
$\varphi_1-\varphi_{2q'+1}\ne 0$, we deduce
\begin{eqnarray}
  \label{eq:IprimeS-2-odd-2-B-x}
\left[\tilde{L}^{-1}_{{2q'}-1}\right]^{(2q')(0)}&=&\left(-\right)^{q'+1}
  \sum_{m'_{2}\ne 0}\sum_{m'_{3}\ne 0} ... \sum_{m'_{{{2q'}}}\ne 0}
  I_{c,1}^{'\left(2q'+1\right)\,\left({2q'}\right)\,\left(1,2,\,...,{{2q'}}\right)}
  \left[-\left(\sum_{\alpha=2}^{2q'}m'_\alpha\right),
    \,m'_2,\,...,\,m'_{{{2q'}}}\right]\times\\ \nonumber &&
  \left(\prod_{\alpha=2}^{{2q'}} m'_{\alpha} \right)
  \cos\left(\left(\sum_{\alpha=2}^{2q'} m'_\alpha\right)
  \left(\varphi_1-\varphi_{2q'+1}\right)\right).
\end{eqnarray}
\end{widetext}

\subsubsection{Summary}

The zero-flux limit of the predicted experimental signal with
$N=2q'+1$ generalizes the above
Eqs.~(\ref{eq:partial1-0})-(\ref{eq:partial1-0-fin}):
\begin{equation}
  {\chi}^{(2q'+1)(0)}\left(\varphi_1-\varphi_{2q}\right)
  =
  \left(\ref{eq:IprimeS-2-odd-2-A}\right)
  +\left(\ref{eq:IprimeS-2-odd-2-B-x}\right)
  .
\end{equation}

\subsection{Discussion}
\label{sec:phys-arg}

To summarize, the $p=N$-terminal and some of the
$p=N-1$-terminal dc-Josephson processes contribute for finite values
to ${\chi}^{\left(N\right)}\equiv\left[L^{-1}_{N-2}\right]'$ given
by Eq.~(\ref{eq:A1}):
\begin{equation}
  \label{eq:x1}
        {\chi}^{\left(N\right)} =\left[L^{-1}_{N-2}\right]'=
        \left[L^{-1}_{N-2}\right]^{'\left(N\right)} +
        \left[L^{-1}_{N-2}\right]^{'(N-1)} .
\end{equation}
In general,
\begin{eqnarray}
  \label{eq:x2-1}
  \left[L^{-1}_{N-2}\right]^{'\left(N\right)} \ne 0\\
  \left[L^{-1}_{N-2}\right]^{'(N-1)} \ne 0
    \label{eq:x2-2}
\end{eqnarray}
if the corresponding critical currents are nonvanishingly small. We find:
\begin{equation}
  \label{eq:x3}
  \left[L^{-1}_{N-2}\right]^{'(p)}=0
  \mbox{ if } p\le N-2
.
\end{equation}

In addition, the $\left(N-2\right)$-th order derivative of
$\sin\left(m_{a'_1} \varphi_{a'_1} + m_{a'_2}\varphi_{a'_2} + .... +
m_{a'_p}\varphi_{a'_p}\right)$ in Eq.~(\ref{eq:A1}) produces $\sin$ or
$\cos$ according to the parity of $N$.

The Appendix details examples with $N=6$ superconducting leads, see
figure~\ref{fig:multiloop}.

\begin{figure*}[htb]
  \begin{minipage}{.66\textwidth}
    \includegraphics[width=\textwidth]{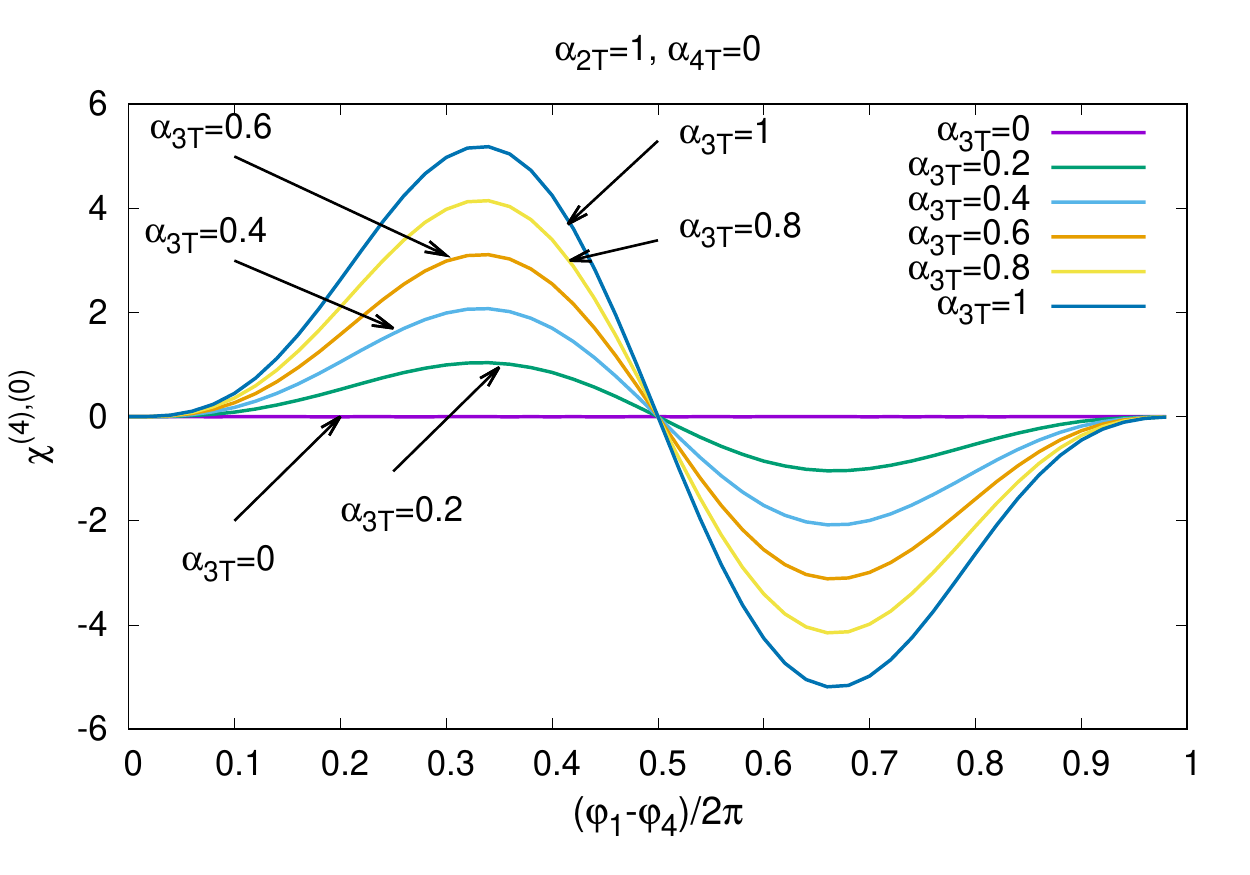}\end{minipage} \begin{minipage}{.33\textwidth} \caption{The
      figure shows $\chi^{(4)(0)}(\varphi_1-\varphi_4)$ as a function
      of $\varphi_1-\varphi_4$, for $\alpha_{2T}=1$,
      $\alpha_{4T}=0$. The values of $\alpha_{3T}$ between $0$ and $1$
      are shown on the figure.
    \label{fig:chi4}}\end{minipage}
\end{figure*}

\section{Numerical results}
\label{sec:num}

Now, we present a numerical illustration that
$\chi^{(4)(0)}(\varphi_1-\varphi_4)$ defined in Eq.~(\ref{eq:S-(4)})
is vanishingly small if the $p=3$ and the $p=4$-terminal critical
currents are also vanishingly small {\it i.e.}
$I_c^{'\left(4\right)\,\left(3\right)\,\left(1,\,3,\,4\right)}=0$ and
$I_c^{'\left(4\right)\,\left(4\right)\,\left(1,\,2,\,3,\,4\right)}=0$
in Eqs.~(\ref{eq:I-panelb-debut-1})-(\ref{eq:star-page-7}).

This implies that ``measurement of $\chi^{(4)(0)}(\varphi_1-\varphi_4)
\ne 0$'' is evidence for the $p=3$ or $p=4$-terminal dc-Josephson
critical currents
$I_c^{'\left(4\right)\,\left(3\right)\,\left(1,\,3,\,4\right)} \ne 0$
or $I_c^{'\left(4\right)\,\left(4\right)\,\left(1,\,2,\,3,\,4\right)}
\ne 0$.

We use the following phenomenological modeling of the
ABS energies:
\begin{widetext}
\begin{eqnarray}
  E_{ABS}(\varphi_1,\varphi_2,\varphi_3,\varphi_4)
  &=&\alpha_0\\ \nonumber &+&
  \alpha_{2T}\left\{\cos(\varphi_1-\varphi_2)+
  \cos(\varphi_1-\varphi_3)+ \cos(\varphi_1-\varphi_4)+
  \cos(\varphi_2-\varphi_3)+ \cos(\varphi_2-\varphi_4)+
  \cos(\varphi_3-\varphi_4)\right\}\\\nonumber &+&
  \alpha_{3T}\left\{\cos(\varphi_1+\varphi_2-2\varphi_3)+
  \cos(\varphi_1+\varphi_2-2\varphi_4)+
  \cos(\varphi_2+\varphi_3-2\varphi_1)+
  \cos(\varphi_2+\varphi_3-2\varphi_4)\right.\\\nonumber &&+\left.
  \cos(\varphi_1+\varphi_3-2\varphi_2)+
  \cos(\varphi_1+\varphi_3-2\varphi_4)+
  \cos(\varphi_1+\varphi_4-2\varphi_2)+
  \cos(\varphi_1+\varphi_4-2\varphi_3)\right.\\\nonumber &&+\left.
  \cos(\varphi_2+\varphi_3-2\varphi_1)+
  \cos(\varphi_2+\varphi_3-2\varphi_4)+
  \cos(\varphi_2+\varphi_4-2\varphi_1)+
  \cos(\varphi_2+\varphi_4-2\varphi_3)\right.  \\\nonumber &&+\left.
  \cos(\varphi_3+\varphi_4-2\varphi_1)+
  \cos(\varphi_3+\varphi_4-2\varphi_2)\right.\\ &+& \alpha_{4T}\left\{
  \cos(\varphi_1-\varphi_2-\varphi_3+\varphi_4)+
  \cos(\varphi_1-\varphi_2+\varphi_3-\varphi_4)+
  \cos(\varphi_1+\varphi_2-\varphi_3-\varphi_4) \right\} ,
\end{eqnarray}
\end{widetext}
where $\alpha_0$ is a constant energy shift, and $\alpha_{2T}$,
$\alpha_{3T}$ and $\alpha_{4T}$ parameterize the $p=2,\,3,\,4$
dc-Josephson effect.

The current $I_1$ through the lead $S_1$ is deduced from
Eq.~(\ref{eq:IS-N0-a1}), and $\chi^{(4)(0)}(\varphi_1-\varphi_4)$ is
obtained from Eqs.~(\ref{eq:S-(4)})-(\ref{eq:S-(4)-bis}).

Figure~\ref{fig:chi4} shows $\chi^{(4)(0)}(\varphi_1-\varphi_4)$ for
$\alpha_{2T}=1$, $\alpha_{4T}=0$ and $\alpha_{3T}$ between $0$ and
$1$. It is concluded that $\alpha_{3T}=\alpha_{4T}=0$ implies
$\chi^{(4)(0)}(\varphi_1-\varphi_4)=0$. Conversely, $\alpha_{3T}\ne 0$
with $\alpha_{4T}=0$ implies nonvanishingly small
$\chi^{(4)(0)}(\varphi_1-\varphi_4)\ne 0$, whatever
$\alpha_{2T}$. Thus, Figure~\ref{fig:chi4} confirms that
$\chi^{(4)(0)}(\varphi_1-\varphi_4) \ne 0$ is evidence for the $p=3$
or $p=4$ dc-Josephson effect in a device with $N=4$ superconducting
leads. {The effect is ``yes or no'' answer to the $p\ge3$
  multiterminal dc-Josephson effect.}

\section{Conclusions}
\label{sec:conclusions}

Now, we present summary of the main results and final remarks.

We considered multiterminal dc-Josephson interferometers that were
investigated in devices containing $N=4$ (or with $N\ge 4$)
superconducting leads, in a way which is different from the previous
Refs.~\onlinecite{Pillet1,Pillet2,Benzoni,Rech}.

Specifically, we proposed to detect the $p\ge3$-terminal Josephson
effect in a device containing $N=4$ superconducting leads, see
figure~\ref{fig:N-terminal-junction}c.  The experimental signal
${\chi}^{(4)}$ or ${\chi}^{(4)(0)}\left(\varphi_1- \varphi_4\right)$
is given by Eqs.~(\ref{eq:S-(4)-gene})-(\ref{eq:S-(4)-gene-bis}) or
Eqs.~(\ref{eq:S-(4)})-(\ref{eq:S-(4)-bis}) respectively. We obtained
the following result:

``{\it Nonvanishingly small second-order nonlocal inverse inductance
  ${\chi}^{(4)}$ or ${\chi}^{(4)(0)}\left(\varphi_1-
  \varphi_4\right)$}''

generically implies

``{\it Evidence for the three or four-terminal dc-Josephson effect}''.

We generalized this statement to a device containing arbitrary number
$N\ge 4$ of the superconducting leads, thus with $N-1$ independent phase
differences. One of those phase differences ({\it e.g.}
$\varphi_1-\varphi_N$) is reserved for controlling the current-phase
relation of the overall two-terminal device. The remaining $N-2$
independent superconducting phase differences are used in the
definition Eq.~(\ref{eq:A1}) of the higher-order nonlocal inverse
inductance ${\chi}^{(N)}$ and they are related to the $N-2$
independent fluxes.

Practically, the partial derivative of the supercurrent $I_1$ through
the superconducting lead $S_1$ with respect to $\varphi_2-\varphi_N$,
$\varphi_3-\varphi_N$, ..., $\varphi_{N-1}-\varphi_N$ [see
  Eqs.~(\ref{eq:S-(4)-gene})-(\ref{eq:S-(4)-gene-bis}),
  (\ref{eq:S-(4)})-(\ref{eq:S-(4)-bis}) and Eq.~(\ref{eq:A1})] can be
evaluated by changing variables from the $N-1$ independent phase
differences $\varphi_1-\varphi_N$, $\varphi_2-\varphi_N$, ...,
$\varphi_{N-1}-\varphi_N$ to $\varphi_1-\varphi_N$ and the $N-2$
fluxes $\Phi_{A_1}$, $\Phi_{A_2}$, ...  , $\Phi_{A_{N-2}}$.

We demonstrated that finite values for ${\chi}^{(N)}\ne 0$ or ${\chi}^{(N)(0)}\left(\varphi_1- \varphi_N\right)\ne 0$ necessarily
originate from the $p=N-1$ and $p=N$-terminal dc-Josephson effects:
the $p$-terminal dc-Josephson effect does not contribute to ${\chi}^{(N)}$ or ${\chi}^{(N)(0)}\left(\varphi_1- \varphi_N\right)$ if
$p\le N-2$.

Finally, a device with number $N=4$ of the superconducting leads and
two loops seems to be already quite interesting from the point of view
of possible experiments, see
figure~\ref{fig:N-terminal-junction}c. Microwave detection of
${\chi}^{(N)}\ne 0$ or ${\chi}^{(N)(0)}\left(\varphi_1-
\varphi_N\right) \ne 0$ is promising, see also the recent
Refs.~\onlinecite{Feinberg1,Feinberg2,topo1-plus-Floquet}.

\section*{Acknowledgements}

The idea of the paper came from reading the manuscript of V. Bezoni's
PhD thesis \cite{Benzoni}. The author wishes to thank
\c{C}.\"{O}. Girit for useful discussions and J.D. Pillet for
sharing their results \cite{Pillet1,Pillet2} prior to making their
preprints public. The author wishes to thank Romain Danneau for useful
comments on a previous version of the manuscript. The author thanks
the Infrastructure de Calcul Intensif et de Donn\'ees (GRICAD) for use
of the resources of the M\'esocentre de Calcul Intensif de
l'Universit\'e Grenoble-Alpes (CIMENT). The author acknowledges
support from the French National Research Agency (ANR) in the
framework of the Graphmon project (ANR-19-CE47-0007).

\appendix

\section*{Appendix: Examples with $N=6$ superconducting leads}
\label{sec:ex}

In this Appendix, we detail the examples shown on
figure~\ref{fig:multiloop} with $N=6$ superconducting leads.

Figure~\ref{fig:multiloop}b shows example of a $p=3$-terminal
dc-Josephson process with $N=6$. Figure~\ref{fig:multiloop}c shows a
$p=6$-terminal process with $N=6$. Figures~\ref{fig:multiloop}d and
~\ref{fig:multiloop}e show $p=5$ with $N=6$. Not all values of $p$
contribute to ${\chi}^{(N)}=\left[L^{-1}_{N-2}\right]'$ in
Eq.~(\ref{eq:A1}):
 
(i) The $p=3$-terminal dc-Josephson current
$i_1(\varphi_3-\varphi_1,\,\varphi_5-\varphi_1)$ shown on
figure~\ref{fig:multiloop}b depends on the two independent phase
differences $\varphi_3-\varphi_1$ and $\varphi_5-\varphi_1$, thus it
cannot contribute to ${\chi}^{(6)}=\left[L^{-1}_{4}\right]'$ in
Eq.~(\ref{eq:A1}) where nonvanishingly small value for
${\chi}^{(6)}=\left[L^{-1}_{4}\right]'\ne 0$ requires additional
sensitivity of the supercurrent through the lead $S_1$ on both the
superconducting phase variables $\varphi_2$ and $\varphi_4$.

(ii) The $p=6$-terminal dc-Josephson current $i_2$ shown on
figure~\ref{fig:multiloop}c is sensitive to all phase differences
between $S_n$ and $S_m$ with $n,\,m=1,\,...,6$, thus it contributes
for a finite value to ${\chi}^{(N)}=\left[L^{-1}_{N-2}\right]'$ in
Eq.~(\ref{eq:A1}), see below.

(iii) The $p=5$-terminal dc-Josephson current
$i_3(\varphi_2-\varphi_1,\,\varphi_4-\varphi_1,
\varphi_5-\varphi_1,\,\varphi_6-\varphi_1)$ shown on
figure~\ref{fig:multiloop}d depends on $N-2=4$ phase
differences. However, this dc-Josephson process produces
${\chi}^{(6)}=\left[L^{-1}_{4}\right]'=0$ because $i_3$ on
figure~\ref{fig:multiloop}d is independent on $\varphi_3$.

(iv) The $p=5$-terminal dc-Josephson current
$i_4(\varphi_2-\varphi_1,\,\varphi_3-\varphi_1,\,\varphi_4-\varphi_1,
\varphi_5-\varphi_1)$ shown on figure~\ref{fig:multiloop}e is
sensitive on all the phase variables
$\{\varphi_1,\,\varphi_2,\,\varphi_3,\,\varphi_4,\,\varphi_5\}$. Then,
$i_4\equiv i'_4(\varphi_1-\varphi_6,\, \varphi_2-\varphi_6,
\,\varphi_3-\varphi_6, \,\varphi_4-\varphi_6, \,\varphi_5-\varphi_6)$
contributes for a finite value to
${\chi}^{(6)}=\left[L^{-1}_{4}\right]'$, see below.

Now, we detail the above items (ii) and (iv) corresponding to
figures~\ref{fig:multiloop}c and~\ref{fig:multiloop}e.

First, concerning item (ii) and figure~\ref{fig:multiloop}c, the
lowest-order supercurrent $i_2$ through the lead $S_1$ is given by
\begin{widetext}
\begin{eqnarray}
  i_2(\varphi_1,\varphi_2,\varphi_3,\varphi_4,\varphi_5,\varphi_6)&=&
  j_2^{(0)} \sin\left[\left(\varphi_2-\varphi_1
  \right)+\left(\varphi_3-\varphi_1\right)+
  \left(\varphi_4-\varphi_1\right)+\left(\varphi_5-\varphi_1\right)
  +\left(\varphi_6-\varphi_1\right)\right]\\ &=& j_2^{(0)}
  \sin\left[-5\varphi_1+\varphi_2+\varphi_3+
  \varphi_4+\varphi_5+\varphi_6\right]\\ &=& j_2^{(0)} \sin\left[-5
  \left(\varphi_1-\varphi_6\right)+\left(
  \varphi_2-\varphi_6\right)
  +\left(\varphi_3-\varphi_6\right)+\left(\varphi_4-\varphi_6\right)
  +\left(\varphi_5-\varphi_6\right)\right]
  .
\end{eqnarray}
Thus,
\begin{eqnarray}
  \label{eq:partial4-1}
  &&
  \frac{\partial^4i_2(\varphi_1,\varphi_2,\varphi_3,\varphi_4,\varphi_5,\varphi_6)}
       {\partial\left(\varphi_2-\varphi_6\right)
         \partial\left(\varphi_3-\varphi_6\right)
         \partial\left(\varphi_4-\varphi_6\right)
         \partial\left(\varphi_5-\varphi_6\right)} =
       i_2(\varphi_1,\varphi_2,\varphi_3,\varphi_4,\varphi_5,\varphi_6)
\end{eqnarray}
and we find:
\begin{eqnarray}
&&\left. \frac{\partial^4
    i_2(\varphi_1,\varphi_2,\varphi_3,\varphi_4,\varphi_5,\varphi_6)}{\partial\left(\varphi_2-\varphi_6\right)
    \partial\left(\varphi_3-\varphi_6\right)
    \partial\left(\varphi_4-\varphi_6\right)
    \partial\left(\varphi_5-\varphi_6\right)} \right|_{\begin{array}{c}
    \varphi_2-\varphi_6=0,\,\varphi_3-\varphi_6=0\\
    \varphi_4-\varphi_6=0,\,\varphi_5-\varphi_6=0
    \end{array}}= -j^{(0)}_2
  \sin\left[5\left(\varphi_1-\varphi_6\right)\right] .
\end{eqnarray}

Second, we consider item (iv) and figure~\ref{fig:multiloop}d:
\begin{eqnarray}
  i_4(\varphi_1,\varphi_2,\varphi_3,\varphi_4,\varphi_5,\varphi_6)
  &=&
  j_4^{(0)}
  \sin\left[\left(\varphi_2-\varphi_1
  \right)+
  \left(\varphi_4-\varphi_1\right)+\left(\varphi_5-\varphi_1\right)
  +\left(\varphi_6-\varphi_1\right)\right]\\
  &=&
  j_4^{(0)}
  \sin\left[-4\varphi_1+\varphi_2+\varphi_4+\varphi_5+\varphi_6\right]\\
   &=&
  j_4^{(0)}
  \sin\left[ -4\left(\varphi_1-\varphi_6\right)
    +\left(\varphi_2-\varphi_6\right)+
    \left(\varphi_4-\varphi_6\right)+\left(\varphi_5-\varphi_6\right)\right]
  ,
\end{eqnarray}
which implies that the following is vanishingly small:
\begin{equation}
  \label{eq:partial4-2}
  \frac{\partial^4
    i_4(\varphi_1,\varphi_2,\varphi_3,\varphi_4,\varphi_5,\varphi_6)}
       {\partial\left(\varphi_2-\varphi_6\right)
         \partial\left(\varphi_3-\varphi_6\right)
         \partial\left(\varphi_4-\varphi_6\right)
         \partial\left(\varphi_5-\varphi_6\right)} 
  =0.
\end{equation}
The above Eqs.~(\ref{eq:partial4-1}) and~(\ref{eq:partial4-2}) confirm the
general theory presented in the paper.

\end{widetext}


\begin{thebibliography}{99}


\bibitem{Josephson} B.D. Josephson, {\it Possible new effects in
  superconductive tunnelling}, Physics Letters {\bf 1}, 251 (1962).

\bibitem{Likharev} K.K. Likharev, {\it Superconducting weak links},
  Rev. Mod. Phys. {\bf 51}, 101 (1979).

\bibitem{Anderson} P.W. Anderson and J.M. Rowell, {\it Probable
  Observation of the Josephson Superconducting Tunneling Effect},
  Phys. Rev. Lett. {\bf 10}, 230 (1963).

\bibitem{theory-CPBS1} J.M. Byers and M.E. Flatt\'e, {\it Probing
  Spatial Correlations with Nanoscale Two-Contact Tunneling},
  Phys. Rev. Lett. 74, 306 (1995).
  
\bibitem{theory-CPBS2} N.K. Allsopp, V.C. Hui, C.J. Lambert, and
  S.J. Robinson, {\it Theory of the sign of multi-probe conductances
    for normal and superconducting materials}, J. Phys.:
  Condens. Matter 6, 10475 (1994).
  
\bibitem{theory-CPBS3} J. Torr\`es and T. Martin, {\it Positive and
  negative Hanbury-Brown and Twiss correlations in normal
  metal-superconducting devices}, Eur. Phys. J. B {\bf 12}, 319
  (1999).
  
\bibitem{theory-CPBS4} M.S. Choi, C. Bruder, and D. Loss, {\it
  Spin-dependent Josephson current through double quantum dots and
  measurement of entangled electron states}, Phys. Rev. B {\bf 62},
  13569 (2000).
  
\bibitem{theory-CPBS5} G. Deutscher and D. Feinberg, {\it Coupling
  superconducting-ferromagnetic point contacts by Andreev
  reflections}, Appl. Phys. Lett. 76, 487 (2000).
  
\bibitem{theory-CPBS6} G. Falci, D. Feinberg, and F.W.J. Hekking,
  {\it Correlated tunneling into a superconductor in a multiprobe
    hybrid structure}, Europhys. Lett. 54, 255 (2001).

\bibitem{theory-CPBS7} R. M\'elin and D. Feinberg, {\it Transport theory
  of multiterminal hybrid structures}, Eur. Phys. J. B 26, 101 (2002).

\bibitem{theory-CPBS8} N.M. Chtchelkatchev, G. Blatter,
  G.B. Lesovik, and T. Martin, {\it Bell inequalities and
    entanglement in solid-state devices}, Phys. Rev. B {\bf 66},
  161320 (2002).
  
\bibitem{theory-CPBS9} R. M\'elin and D. Feinberg, {\it Sign of the
  crossed conductances at a ferromagnet/superconductor/ferromagnet
  double interface}, Phys. Rev. B 70, 174509 (2004).

\bibitem{exp-CPBS1} D. Beckmann, H.B. Weber, and H.v. Löhneysen, {\it Evidence for crossed Andreev reflection in Superconductor-Ferromagnet hybrid structures}, Phys. Rev. Lett. {\bf 93}, 197003 (2004).

\bibitem{exp-CPBS2} S. Russo, M. Kroug, T.M. Klapwijk, and
  A.F. Morpurgo, {\it Experimental observation of bias-dependent
    nonlocal Andreev reflection}, Phys. Rev. Lett. {\bf 95}, 027002
  (2005).

\bibitem{exp-CPBS3} P. Cadden-Zimansky and V. Chandrasekhar, {\it
  Nonlocal correlations in normal-metal superconducting systems},
  Phys. Rev. Lett. {\bf 97}, 237003 (2006).

\bibitem{exp-CPBS4} P. Cadden-Zimansky, Z. Jiang, and V. Chandrasekhar,
  {\it Charge imbalance, crossed Andreev reflection and elastic
    co-tunnelling in ferromagnet/superconductor/normal-metal
    structures}, New J. Phys. {\bf 9}, 116 (2007).

\bibitem{exp-CPBS5} L.G. Herrmann, F. Portier, P. Roche, A. Levy
  Yeyati, T. Kontos, and C. Strunk, {\it Carbon nanotubes as Cooper
    pair beam splitters}, Phys. Rev. Lett. {\bf 104}, 026801 (2010).
  
\bibitem{exp-CPBS6} L. Hofstetter, S. Csonka, J. Nygoard, and
  C. Sch\"onenberger, {\it Cooper pair splitter realized in a
    two-quantum-dot Y-junction}, Nature (London) {\bf 461}, 960
  (2009).

\bibitem{exp-CPBS7} J. Wei and V. Chandrasekhar, {\it Positive noise
  cross-correlation in hybrid superconducting and normal-metal
  three-terminal devices}, Nat. Phys. 6, 494 (2010).
  
\bibitem{exp-CPBS8} A. Das, Y. Ronen, M. Heiblum, D. Mahalu,
  A.V. Kretinin, and H.  Shtrikman, {\it High-efficiency Cooper pair
    splitting demonstrated by two-particle conductance resonance and
    positive noise cross- correlation}, Nat. Commun. {\bf 3}, 1165
  (2012).

\bibitem{Freyn} A. Freyn, B. Dou\c{c}ot, D. Feinberg and
  R. M\'elin, {\it Production of non-local quartets and
    phase-sensitive entanglement in a superconducting beam splitter},
  Phys. Rev. Lett. {\bf 106}, 257005 (2011).

\bibitem{Melin-EPJB} R. M\'elin, D. Feinberg and
  B. Dou\c{c}ot, {\it Partially resummed perturbation theory for
    multiple Andreev reflections in a short three-terminal Josephson
    junction}, Eur. Phys. J. B {\bf 89}, 67 (2016).

\bibitem{QUARTETS1} T. Jonckheere, J. Rech, T. Martin, B. Dou\c{c}ot,
  D. Feinberg, and R. M\'{e}lin, \textit{Multipair DC Josephson
    resonances in a biased allsuperconducting bijunction},
  Phys. Rev. B \textbf{87}, 214501 (2013).

\bibitem{Rech} J. Rech, T. Jonckheere, T. Martin, B. Dou\c{c}ot,
  D. Feinberg, and R. M\'elin, \textit{Proposal for the observation of
    nonlocal multipair production}, Phys. Rev. B \textbf{90}, 075419
  (2014).
    
\bibitem{QUARTETS2} R. M\'elin, M. Sotto, D. Feinberg, J.-G. Caputo and
  B. Dou\c{c}ot, {\it Gate-tunable zero-frequency current
    cross-correlations of the quartet mode in a voltage-biased
    three-terminal Josephson junction}, Phys. Rev. B {\bf 93}, 115436
  (2016).

\bibitem{QUARTETS3} R. M\'elin, J.-G. Caputo, K. Yang and B. Dou\c{c}ot,
  {\it Simple Floquet-Wannier-Stark-Andreev viewpoint and emergence of
    low-energy scales in a voltage-biased three-terminal Josephson
    junction}, Phys. Rev. B {\bf 95}, 085415 (2017).

\bibitem{QUARTETS4} R. M\'elin, R. Danneau,
  K. Yang, J.-G. Caputo, and B. Dou\c{c}ot, {\it Engineering the
    Floquet spectrum of superconducting multiterminal quantum dots},
  Phys. Rev. B {\bf 100}, 035450 (2019).

\bibitem{QUARTETS5} B. Dou\c{c}ot, R. Danneau, K. Yang, J.-G. Caputo and
  R. M\'elin, {\it Berry phase in superconducting multiterminal
    quantum dots}, Phys. Rev. B {\bf 101}, 035411 (2020).

\bibitem{QUARTETS6} R. M\'elin, {\it Inversion in a four-terminal
  superconducting device on the quartet line: I. Two-dimensional metal
  and the quartet beam splitter}, Phys. Rev. B {\bf 102}, 245435
  (2020).

\bibitem{QUARTETS7} R. M\'elin and B. Dou\c{c}ot, {\it Inversion in a
  four terminal superconducting device on the quartet line:
  II. Quantum dot and Floquet theory}, Phys. Rev. B {\bf 102}, 245436
  (2020).

\bibitem{Lefloch} A.H. Pfeffer, J.E. Duvauchelle, H. Courtois,
  R. M\'elin, D. Feinberg, and F. Lefloch, {\it Subgap structure in the
    conductance of a three-terminal Josephson junction}, Phys. Rev. B
  {\bf 90}, 075401 (2014).
 
\bibitem{Heiblum} Y. Cohen, Y. Ronen, J.H. Kang, M. Heiblum,
  D. Feinberg, R.  M\'elin and H. Strikman, {\it Non-local
    supercurrent of quartets in a three-terminal Josephson junction},
  Proc. Natl. Acad. Sci. U. S. A. \textbf{115}, 6991 (2018).

\bibitem{Kim} K.F. Huang, Y. Ronen, R. M\'elin,
  D. Feinberg, K. Watanabe, T. Taniguchi and P. Kim, {\it Quartet
    supercurrent in a multi-terminal Graphene-based Josephson
    Junction}, cond-mat preprint (2020).

\bibitem{multiterminal-exp1}  E. Strambini, S. D'Ambrosio, F. Vischi,
  F.S. Bergeret, Yu.V. Nazarov, and F. Giazotto, \textit{The
    $\omega$-SQUIPT as a tool to phase-engineer Josephson topological
    materials}, Nat. Nanotechnol. \textbf{11}, 1055 (2016).
  
\bibitem{multiterminal-exp2} A.W. Draelos, M.-T. Wei, A. Seredinski, H. Li,
  Y. Mehta, K. Watanabe, T. Taniguchi, I.V. Borzenets, F. Amet, and
  G. Finkelstein, \textit{Supercurrent flow in multiterminal graphene
    Josephson junctions}, Nano Lett. \textbf{19}, 1039 (2019).  
  
\bibitem{multiterminal-exp3} N.  Pankratova, H.  Lee, R. Kuzmin, K.
  Wickramasinghe,1, W. Mayer,J. Yuan,M. Vavilov,J. Shabani and
  V. Manucharyan, {\it The multi-terminal Josephson effect},
  Phys. Rev. X {\bf 10}, 031051 (2020).

\bibitem{multiterminal-exp4} G.V. Graziano, J.S. Lee, M. Pendharkar,
  C. Palmstrom and V.S. Pribiag, {\it Transport Studies in a
    Gate-Tunable Three-Terminal Josephson Junction},
  arXiv:1905.11730v2 (2020).
  
\bibitem{multiterminal-exp5} E.G. Arnault, T. Larson, A. Seredinski,
  L. Zhao, H. Li, K. Watanabe, T. Tanniguchi, I. Borzenets, F. Amet
  and G. Finkelstein, {\it The multiterminal inverse AC Josephson
    effect}, arXiv:2012.15253v1 (2020).

\bibitem{multiterminal-exp6} S.A. Khan, L. Stampfer, T. Mutas,
  J.-H. Kang, P. Krogstrup and T.S. Jespersen, {\it Multiterminal
    Quantized Conductance in InSb Nanocrosses}, arXiv:2101.02529
  (2021).
  
\bibitem{Nazarov1} R.-P. Riwar, M. Houzet, J.S. Meyer, and
  Y.V. Nazarov, \textit{Multi-terminal Josephson junctions as
    topological materials}, Nat. Commun. 7, 11167 (2016).

\bibitem{Nazarov2} E. Eriksson, R.-P. Riwar, M. Houzet, J. S. Meyer,
  and Y. V. Nazarov, {\it Topological transconductance quantization in
    a four-terminal Josephson junction}, Phys. Rev. B {\bf 95}, 075417
  (2017).
  
\bibitem{topo0} O. Deb, K. Sengupta and D. Sen, {\it Josephson
  junctions of multiple superconducting wires}, Phys. Rev. B 97,
  174518 (2018).
  
\bibitem{topo1} H. Weisbrich, R.L. Klees, G. Rastelli and W. Belzig,
  {\it Second Chern Number and Non-Abelian Berry Phase in Topological
    Superconducting Systems}, PRX Quantum {\bf 2}, 010310 (2021).

\bibitem{topo2} V. Fatem, A.R. Akhmerov and L. Bretheau,
  {\it Weyl Josepshon circuits}, arXiv:2008.13758v1 (2020).

\bibitem{topo3} L. Peyruchat, J. Griesmar, J.-D. Pillet and \c{C}.\"O
  Girit, {\it Transconductance quantization in a topological Josephson
    tunnel junction circuit}, arXiv:2009.03291v1 (2020).

\bibitem{topo4} Y. Chen and Y.V. Nazarov, {\it Weyl point immersed in
  a continuous spectrum: an example from superconducting
  nanostructures}, arXiv:2102.03947v1 (2021).

\bibitem{topo5} E.V. Repin and Y.V. Nazarov, {\it Weyl points in the
  multi-terminal Hybrid Superconductor-Semiconductor Nanowire devices},
  arXiv:2010.11494v1 (2020).

\bibitem{Levchenko1} H.-Y. Xie, M.G. Vavilov and A. Levchenko, {\it
  Topological Andreev bands in three-terminal Josephson junctions},
  Phys. Rev. B {\bf 96}, 161406 (2017).

\bibitem{Levchenko2} H.-Y. Xie, M.G. Vavilov and A. Levchenko, {\it
  Weyl nodes in Andreev spectra of multiterminal Josephson junctions:
  Chern numbers, conductances and supercurrents}, Phys. Rev. B {\bf
  97}, 035443 (2018).
  
\bibitem{Berry} B. Dou\c{c}ot, R. Danneau, K. Yang, J.-G. Caputo and
  R. M\'elin, {\it Berry phase in superconducting multiterminal
    quantum dots}, Phys. Rev. B {\bf 101}, 035411 (2020).
  
\bibitem{Feinberg1} B. Venitucci, D. Feinberg, R. M\'elin,
  B. Dou\c{c}ot, {\it Nonadiabatic Josephson current pumping by
    microwave irradiation}, Phys. Rev. B {\bf 97}, 195423 (2018).

\bibitem{Feinberg2} L.P. Gavensky, G. Usaj, D. Feinberg and
  C.A. Balseiro, {\it Berry curvature tomography and realization of
    topological Haldane model in driven three-terminal Josephson
    junctions}, Phys. Rev. B {\bf 97}, 220505 (2018).
  
\bibitem{topo1-plus-Floquet} R. L. Klees, G. Rastelli, J. C. Cuevas, and
  W. Belzig, {\it Microwave Spectroscopy Reveals the Quantum Geometric
    Tensor of Topological Josephson Matter}, Phys. Rev. Lett. {\bf
    124}, 197002 (2020).
  
\bibitem{Pillet1} J.D. Pillet, V. Benzoni, J. Griesmar, J.-L. Smirr and
  \c{C}. \"{O}. Girit, {\it Nonlocal Josephson Effect in Andreev
    Molecules} Nano Lett. {\bf 19}, 7138 (2019).

\bibitem{Pillet2} J.-D. Pillet, V. Benzoni, J. Griesmar, J.-L. Smirr
  and \c{C} \"O Girit, {\it Scattering description of Andreev
    molecules}, SciPost Phys. Core {\bf 2}, 009 (2020).
  
\bibitem{Benzoni} V. Benzoni, {\it Hybridization of Andreev bound
  states in closely spaced Josephson junctions}, PhD thesis, Th\`ese
  de Doctorat de l'Universit\'e PSL (2021).

\bibitem{SQUID} J.E. Zimmerman and A.H. Silver, {\it Macroscopic
  Quantum Interference Effects through Superconducting Point
  Contacts}, Phys. Rev. {\bf 141}, 367 (1966).

\bibitem{Zazunov} A. Zazunov, V.S. Shumeiko, E.N. Bratus’,
    J. Lantz, and G. Wendin, {\it Andreev Level Qubit},
    Phys. Rev. Lett. {\bf 90}, 087003 (2003).

\bibitem{Meng} T. Meng, S. Florens and P. Simon, {\it Self-consistent
  description of Andreev bound states in Josephson quantum dot
  devices}, Phys. Rev. B {\bf 79}, 224521 (2009).
  
     
\end{thebibliography}
\end{document}